\documentclass[a4paper, 11pt]{article}
\usepackage{float}
\date{March  2009}
\usepackage{epsfig}
\newcommand{\be}{\begin{equation}}
\newcommand{\ee}{\end{equation}}
\newcommand{\ba}{\begin{eqnarray}}
\newcommand{\ea}{\end{eqnarray}}
\newcommand{\bi}{\begin{itemize}}
\newcommand{\ei}{\end{itemize}}

\newcommand{\<}{\langle}
\renewcommand{\>}{\rangle}

\newcommand{\la}{\label}

\def\a{\alpha}      \def\b{\beta}         \def\G{\Gamma}
\def\d{\delta}        \def\e{\varepsilon}
          \def\l{\lambda}     \def\L{\Lambda}
\def\m{\mu}                     
\def\n{\nu}             
     \def\s{\sigma}  
\def\t{\tau}          
              
\def\w{\omega}        
  \def\OO{{\cal O}}

\newcommand\Tr{\mbox{Tr}\, }
\def\sumint{\hbox{$\sum$}\!\!\!\!\!\!\int}
\def\VV{\cal V}

\begin{document}

\begin{titlepage}
\begin{flushright}
RBRC 1025\,,\,\,NIKHEF 2013-017
\end{flushright}

\begin{centering}
\vfill

 \vspace*{2.0cm}
{\bf \Large Two-loop perturbative corrections to the thermal effective potential
  in gluodynamics}

\vspace{2.0cm}
\centerline{{\bf Adrian Dumitru}\footnote{dumitru@quark.phy.bnl.gov}}
\centerline{RIKEN BNL Research Center, Brookhaven National
  Laboratory, Upton, New York 11973, USA}
\centerline{Department of Natural Sciences, Baruch College, CUNY,}
\centerline {17 Lexington Avenue, New York, New York 10010, USA}
\centerline{The Graduate School and University Center, The City
  University of New York,}
\centerline {365 Fifth Avenue, New York, New York 10016, USA}

\vspace{0.5cm}

\centerline{\bf Yun Guo\footnote{yun.guo@usc.es}}
\centerline{Departamento de F\'isica de Part\'iculas, Universidade de Santiago de Compostela,}
\centerline{E-15782 Santiago de Compostela, Galicia, Spain}
\centerline{Physics Department, Guangxi Normal University, 541004 Guilin, China}

\vspace{0.5cm}
\centerline{\bf Chris P.\ Korthals Altes\footnote{altes@cpt.univ-mrs.fr}}
\centerline{Centre Physique Th\'eorique, au CNRS Case 907,}
\centerline{Campus de Luminy, F-13288 Marseille, France}
\centerline{NIKHEF theory group, Science Park 105, 1098 XG Amsterdam, The Netherlands}

\vspace*{2.0cm}

\end{centering}
\vfill
\centerline{\bf Abstract }
\vspace{0.1cm}
\noindent
The thermodynamics of pure glue theories can be described in terms of
an effective action for the Polyakov loop. This effective action is of
the Landau-Ginzburg type and its variables are the angles
parametrizing the loop. In this paper we compute perturbative
corrections to this action. Remarkably, two-loop corrections turn out
to be proportional to the one-loop action, independent of the
eigenvalues of the loop. By a straightforward generalization of the 't~Hooft
coupling this surprisingly simple result holds for any of the
classical and exceptional groups.

\vfill

 \end{titlepage}

\setcounter{footnote}{0}

\section{INTRODUCTION}\la{ intro}

Understanding of the deconfined phase QCD at high temperature $T$ is
growing, as a result of the heavy ion experiments at RHIC and the LHC, and
theoretical and numerical lattice work. The latter, in particular, has
shown that the conformal anomaly, which is the energy density minus
three times the pressure, in the deconfined phase of $SU(N)$ pure
gauge theories is approximately proportional to $T^2$, up to
temperatures several times the critical temperature $T_c$ for
deconfinement~\cite{e-3p_data}. This demonstrates that $e-3p$ is not
dominated by a constant ``bag pressure.'' Even without a detailed
understanding of the physical origin of this behavior, previous work
has shown that it can be parametrized as a dimension-two constant times
a condensate for the {\em eigenvalues} of the Polyakov loop which
``evaporates'' at high
$T$~\cite{e-3p_cond,Dumitru:2010mj,Dumitru:2012fw}. At very high
temperature then, the behavior of $(e-3p)/T^2$ for $SU(3)$ is described
very well by ``hard thermal loop'' resummed perturbation
theory~\cite{Andersen:2011ug}.

The goal of the present paper is to compute perturbative quantum
corrections to the pressure of that classical condensate at two-loop
order. That is, we compute the leading correction due to interactions
among gluons in the presence of the condensate. Our main tool is the
effective action as a function of the {\it eigenvalues} of the
Polyakov loop. While the two-loop correction to the effective
potential does not affect the interaction measure,\footnote{$\partial
  g^2(T)/\partial T = {\cal O}(g^4)$ is beyond the order considered
  here.} it does of course modify the pressure and the energy density
of the gluon plasma. We hope that our results may be useful for
improving the
models~\cite{e-3p_cond,Dumitru:2010mj,Dumitru:2012fw,Sasaki:2012bi},
which typically employ the one-loop effective potential. Furthermore,
our efforts show that the models mentioned above can be improved
systematically, at least in regard to the perturbative component,
rather than offering mere parametrizations of the lattice results.

We study hot gluodynamics for any number of colors, and make use of
the global $Z(N)$ symmetry~\cite{thooft78} in that system. However,
lattice simulations for groups without a
center~\cite{Greensite:2006sm, mpanero,Wellegehausen:2010ta} show that deconfinement
does not require a global symmetry.  Hence, aside from $SU(N)$ we
perform our perturbative two-loop calculations also for all other
classical gauge groups, including the exceptional group $G(2)$.

For $SU(N)$ gluodynamics, there is an order parameter associated with
the global symmetry $Z(N)$, i.e.  the Polyakov loop,
\be {\bf L}(\vec
x)={\cal P}\,
\exp\left( i\int_0^{1/T}d\t A_0(\vec x,\t) \right) \la{ploop}\,.
\ee
The global symmetry acts on the loop as a large gauge transformation
$\Omega_k(\vec x,\t)$. It develops a $Z(N)$ valued discontinuity
$\exp(ik{2\pi\over N})$ when we move in the periodic Euclidean time
direction $\t$.

The loop ${\bf L}(\vec x)$ transforms like an adjoint field and hence
its trace,
\be
{1\over N}\Tr{\bf L}(\vec x)\,,
\ee
picks up a $Z(N)$ phase,
\be
{1\over N}\Tr{\bf L}(\vec x) \rightarrow \exp\left( ik{2\pi\over N}\right)
{1\over N}\Tr{\bf L}(\vec x)\,. \la{zntransfloop}
\ee

The effective action is simply the traditional path integral over the
gauge fields subject to a
constraint~\cite{O'Raifeartaigh:1986hi}. This constraint is obviously
that the integration is done while preserving the value of the
Polyakov loop at some fixed value $\ell$. Doing so, one generates a probability
distribution for the eigenvalues determined by $\ell$. We are
interested in a loop which is constant in space, so the constraint
should be a delta function with the argument
\be
\ell-{1\over N}\Tr\overline{ {\bf L}}\,,
\la{constraint1}
\ee
involving the spatially averaged loop,
\be
\Tr\overline{{\bf L}}={1\over V}\int_V d\vec x \, \Tr {\bf L}(\vec x)\,.
\ee
Clearly, to fix all independent phases $\Phi$, one has to take as many
powers of the loop as there are independent phases. To avoid clutter
we do not write these higher powers explicitly; in this simplified
notation the effective action becomes
\be \exp(-V{\cal V}(\ell))=\int DA_\m\, \d(
\ell-{1\over N}\Tr\overline{ {\bf L}})\, \exp(-\frac{1}{g^2}S(A)) \la{effpot}\,.
\ee
Hence, for $SU(2)$, where there is only one independent phase, this
expression fully fixes the phase of the loop. So, there is
just one real constraint on the full path integral. We take $\ell$ to
be a trace over a diagonal matrix, without loss of generality (see
Sec. \ref{sec:general}). The loop ${\bf\overline L} $ is, in
general, not diagonal because the fluctuating scalar potential $A_0$
is arbitrary so long as it satisfies the constraint. As we will see,
in perturbation theory the constraint amounts to taking out the $N-1$
zero modes of the fluctuation matrix, for $SU(N)$.

The constraint is nonlinear in the fluctuations, since the Polyakov
loop is so. As consequence, at two and higher loop order there is an
extra vertex involving the zero mode. It generates diagrams with
radiative corrections inserted into the Polyakov
loop~\cite{Belyaev:1991gh}~\cite{KorthalsAltes:1993ca}. They are
crucial for gauge invariance of the effective potential, and we
will use them in this paper.

To compute ${\cal V}(\ell)$ at small coupling fluctuations around the
background of a constant Polyakov loop, $\ell$ are integrated
over. This leads to the gluon black body radiation contribution plus a
$Z(N)$ invariant polynomial of fourth order in the phases $\Phi$ of
the loop. To be specific, we consider the $SU(2)$ gauge group. The loop
has only one independent phase, $2q_1=-2q_2=q$. In terms of the
variable $q$, one obtains~\cite{Gross:1980br,Weiss:1980rj}
\be
{\cal V}_{pert}(q) = -{\pi^2\over{15}}T^4+{4\pi^2\over 3}T^4\;
q^2(1-q)^2\,.    \la{su2pot}
\ee
In this expression, $q$ is defined modulo 1 and a $Z(2)$
transformation corresponds to $q\to 1-q$. The minima of this $Z(2)$
invariant polynomial are at $0$ and $1$, where the loop $\ell=\pm
1$. The motivation of this paper is to establish how radiative
corrections affect this potential. An earlier answer to this
question~\cite{Gocksch:1993iy} presented an elegant but formal proof
using the Vafa-Witten trick~\cite{Vafa:1984xg}, bypassing issues
related to infrared divergences.

The implication of our work is that
\bi
\item {\em in perturbation theory} the eigenvalue distribution
  $\sim\exp(-V{\cal V}_{pert})$ is not affected by two-loop
  corrections.\footnote{However, it may change at higher orders.} In
  particular, the expectation value of the Polyakov loop calculated at
  these minima remains $\ell=\pm 1$.
\item the pressure calculated from the minimum of the potential
  equals the known perturbative pressure calculated at $q=0$.
\ei
However, we stress that when a nonperturbative contribution is
added~\cite{Dumitru:2010mj,Dumitru:2012fw}, that the two-loop
corrections to the perturbative potential do modify the total
result. We postpone a detailed fit to lattice results to a future
publication.

The bullet points above are corroborated by the two loop contribution
to the perturbative potential which we compute explicitly in
Sec.~\ref{sec:twoloop}.  Sec.~\ref{sec:general} contains a
discussion of simple properties of the effective action and the gauge independence of
our corrections. In Sec.~\ref{sec:twoloop} we discuss the
explicit result at two loops; the simplified expressions for the insertion diagram are given in
Sec.~\ref{sim}; the last section contains the conclusions.

\section{GENERALITIES OF THE EFFECTIVE POTENTIAL}\la{sec:general}

For the sake of notation and clarity we will mostly work in this
section with the $SU(2)$ gauge group. In the first subsection we
discuss the relation between various ways of defining the effective
action.  Its expansion about a constant Polyakov loop background is
analyzed next, and finally extended Becchi-Rouet-Stora(BRS) identities are derived which
give us a very useful control over the perturbative expansion.

\subsection{Two ways of obtaining the effective action}
First we introduce a definition of the effective potential which is
manifestly gauge invariant, manageable on the lattice
and, most
importantly for this paper, has relatively simple Feynman rules. It is
the constrained effective action, defined in a large three volume
$V$ in Euclidean space, and periodic in Euclidean time direction with
period $1/T$.

The Polyakov loop was defined above as
\be
{\bf L}(\vec x)={\cal P}\; \exp\left(i\int _0^{1/T}A_0(\vec x,\t)d\t\right)\,.
\la{polloop}
\ee
%
The effective potential is~\cite{O'Raifeartaigh:1986hi}
\be
\exp(-V{\cal V}(\ell))\equiv\int DA_\m\,
\d(\ell-{1\over 2}\Tr{\overline{\bf L}})\,
\exp(-{1\over{g^2}}S(A))\,.
\la{defeffpot}
\ee
Here $\ell$ is some a {\em priori} specified number.

The partition function $Z$ equals
\ba
Z&\equiv& \int DA_\m\exp(-{1\over{g^2}}S(A))\,,\\
\Tr\overline{ {\bf L}}&\equiv&{1\over V}\int_V d\vec x \, \Tr {\bf L}(\vec x)\,.
\ea
The integration is over fields which are periodic in the Euclidean
time direction with period $1/T$.  Note that the $SU(2)$ matrix ${\bf
  L}(\vec x)$ can be diagonalized at every point $\vec x$ by a gauge
transformation. Hence the space-averaged trace of the loop becomes a
spatial average over a cosine, a number not larger than 1. If all the
eigenphases of ${\bf L}(\vec x)$ are aligned, the space average, of
course, becomes the cosine of the common eigenphase.

Thus $\ell$ is bound by\footnote{Modulo renormalization effects, see
e.g.~\cite{Dumitru:2003hp}.}
\be
-1\le \ell\le 1\,,
\ee
and can be parametrized as the trace of a constant SU(2) matrix,
\be
\ell({\bf q})={1\over 2}\Tr\exp(2\pi i {\bf q})\,,
\la{ellasfatq}
\ee
with
\be
{\bf q}=\pmatrix{q_1&0\cr 0&q_2\cr}\,,~~~q_1=-q_2\,.
\ee
Now we will show that in the large volume limit the
definition~(\ref{defeffpot}) is equivalent to the traditional
definition of the effective potential where a
source term,\footnote{We absorb the $1/N$ normalization of the trace of
${\bf L}$ into the source $j$.}
\be
jV\, \Tr{\overline{\bf L}}=j\int_V d\vec x~\Tr {\bf
  L}(\vec x)\,,
\ee
 is introduced into the path integral $Z$,
\ba \exp(-VW(j))=\int
DA_\m\exp\left(-{1\over{g^2}}S(A)-j\int_V d\vec x~\Tr {\bf L}(\vec x)\right)\,,
\ea
with
\ba
\ell&\equiv&\<\Tr\overline{{\bf L}}\>={\partial
  W\over{\partial j}}\,,\nonumber\\
G(\ell)&\equiv& W(j)-j\ell\,.
\ea
The effective action $G(\ell)$ depends on the source $j$ only through
$\ell$, and it satisfies
\be {\partial G\over{\partial{\ell}}}=-j\,.
\la{effactionsource}
\ee
To compare this definition of the effective action $G(\ell)$ to the
one in Eq.~(\ref{defeffpot}), we Laplace transform the latter with $\int
d\ell \exp(-Vj\ell)$,
\be \int d\ell \,
\exp(-Vj\ell)\; \exp(-V{\cal V}(\ell))=\exp(-VW(j))~.  \la{laplacew}
\ee
Steepest descent of this integral in the large $V$ limit tells us that
the effective potential obeys
\ba {\partial{\cal V}(\ell)\over{\partial {\ell}}}&=&-j\,,\nonumber\\ {\cal
  V}(\ell)+j\ell&=&W(j)\,,
\ea
and so
\be G(\ell)={\cal V}(\ell)+f(T)\,.
\ee
This means that both definitions give the same effective potential, up to
a temperature dependent but $\ell$-independent function. However, the
constrained version is well adapted to lattice calculations and admits
a straightforward saddle point expansion around any value of
$\ell$. This expansion is discussed in the next two sections.

The source term $j\int_V d\vec x~\Tr{\bf L}(\vec x)$ is manifestly
gauge invariant. Hence, we expect the effective potential to be the same
in whatever gauge we calculate it. This will turn out to be true
although the expectation value of the space-averaged loop
$\<\Tr{\overline {\bf L}}\>$ is gauge dependent (except at
$j=0$). This gauge dependence precisely cancels that of the free-energy contributions as we will expound in Secs. (\ref{subsec:insertion}) and (\ref{subsec:brs}).

After Fourier transforming the delta function constraint, we
can write the constrained path integral as
\ba
\exp(-V{\cal V}(\ell))&\equiv&\int DA_\m
~d\e\, \exp\bigg(-{1\over{g^2}}S_{con}(A,\e)\bigg)\,,\nonumber\\
S_{con}(A,\e) &=& i{\e}
\left(\ell-{1\over 2}\Tr{\overline{\bf L}}\right) + 
S(A)\,.
\la{epspathint}
\ea
We have traded the constraint for an extra field $\e$ in the path
integral, and we added a phase to the original gauge action. The new
field $\e$ is therefore gauge invariant, like the constraint it
generates.

\subsection{The effective potential in perturbation theory}\la{sec:include}
Below we shall give explicitly the Feynman rules for
fluctuations around a fixed background $\e_c$ and $B_\m$. The
background is supposed to be a minimum of $S_{con}$.
A simple choice of background is
\be
\e_c=0, ~A_\m=B\d_{\m,0}, ~B~\mbox{ constant in space time}.
\la{minimum}
\ee
This is indeed an extremum of $S_{con}$ when minimizing over
the fluctuations in
\be
A_\m=B\d_{\m,0}+gQ_\m, ~\mbox{and}~\e=\e_c+g\e_q.
\la{fluctuations}
\ee
%

%

The gauge zero modes have to be tamed by
introducing gauge-fixing and ghost terms,
\be
S_{gauge}=S(A)+S_{gf}+S_{gh}=S_{free}+gS_{int}\,.
\ee
Our choice of gauge fixing is covariant background gauge,
\be
S_{gf}={1\over {\xi}}\int d\vec x ~ d\t \Tr(D_{\m}(B)Q_\m)^2\,,
~~D_\m(B)=\partial_\m+i[B\d_{\m,0}\,,\cdots],
\la{gaugefix}
\ee
and the constrained action $S_{con}$ changes accordingly into
\be
S_{con}=i\e\left(\ell-{1\over 2}\Tr{\overline{\bf L}}\right) +S_{gauge}.
\la{Scongaugefix}
\ee

Expand $S_{con}(A,\e)$ in terms of $Q$ and $\e_q$ as
\be
S_{con}(B+gQ,\e_c+g\e_q)=\sum_{n\ge 0}^{\infty}g^{n}S_{con}^{(n)}\,.
\la{sconexpand}
\ee
To avoid clutter in the formulas we shall use the following notation
for the expansion in powers of the fluctuation field $Q_0$:
\be {\bf L}(B+gQ_0)={\bf L}(B)+gQ_0\cdot{\bf L}'(B) +g^2Q_0^2\cdot
    {\bf L}''(B)+\OO(g^3Q_0^3)\,,
\ee
and similar for the action $S_{gauge}$. The operator `` $\cdot$ " means integration
over space time and summation over (Lorentz and) color indices.

One mode becomes particularly important in this expansion. It is the
zero Matsubara frequency of the zero momentum mode $\overline
Q_0(\t)=\int_V d\vec x\, Q_0(\vec x,\t)/V$ 
\be
\overline{\overline{Q}}_0\equiv \int_0^{1/T} d\t\, \overline{Q}_0(\t)\,
\ee
It is the {\it only} mode that can produce linear terms in the
expansion around the space time independent background. All other
modes $Q_\m(\vec x,\t)$ are orthogonal to $B$.

The terms linear in the fluctuations [i.e. in $S_{con}^{(1)}({\bf
  q})$] are required to vanish. They are
\ba
i\e_c\Tr\bigg(\overline{\overline{Q}}_0{\bf L}'(B)\bigg)+\Tr\bigg(\overline{\overline{Q}}_0S'_{guage}(B)\bigg)=0\,,\nonumber \\
\e_q(\ell-{1\over 2}\Tr{\bf L}(B))=0\,.
\ea
From the first condition it follows that $\e_c=0$, because $S'_{guage}(B)=0$.
The second fixes $B$ in terms of $\ell$,
\be
\ell-{1\over 2}\Tr{\bf L}(B)=0\,.
\ee
Hence, from~(\ref{ellasfatq}) we have
the background $B$ fixed in terms of the phases $q_1$
and $q_2=-q_1$ of the Polyakov loop,
\be
B=2\pi{\bf q}T\,.
\la{Btoq}
\ee
Hence $S^{(0)}_{con}({\bf q})=0$.

In what follows we write the zero momentum and zero Matsubara
frequency  mode   projected onto ${\bf L}'({\bf q})$ as
\be
\Tr\bigg(\overline{\overline{Q}}_0{\bf L}'({\bf q})\bigg)\equiv\widehat Q_0\,.
\la{hatzeromom}
\ee
The quadratic term in the expansion of $S_{con}$ is therefore the
first nonvanishing term,
\be
S^{(2)}_{con}=\int d\vec x d\t \bigg(\Tr Q_\m(\vec x,\t)(-D^2({\bf q}))\d_{\m\n}+(1-\xi)D_\m({\bf q})D_\n({\bf q})Q_\n(\vec x,\t)\bigg)-i\e_q\widehat Q_0,
\la{s2explicit}
\ee
where we wrote the explicit form of $S''_{gauge}$.

Thus the expansion of the effective potential ($\eta$ and $\w$ are the
ghost fields) becomes
\be
\exp(-V{\VV})=\int DQ_\m D\bar\eta D\w d\e_q\exp(-Q^2\cdot S''_{gauge}({\bf q})
           -i\e_q\widehat Q_0)(1-R)\,. \la{expansioneff}
\ee
The last factor equals
\ba
 1-R&=&1-g Q^3\cdot S'''_{gauge}({\bf q}) - g^2 Q^4\cdot S''''_{gauge}({\bf q}) \nonumber\\
   &+&g^2 i\e_q\Tr \bigg(\overline{Q_0^2}\cdot{\bf L}''({\bf
   q})\bigg)Q^3\cdot S'''_{gauge}({\bf q})+ \cdots\,.
\la{expansioninteract}
\ea
Let us discuss~(\ref{expansioneff}) and~(\ref{expansioninteract}).  The terms
in the exponent are familiar, except for the last one.  This
latter term, after integration over $\e_q$, is restoring the delta
function constraint. It tells us {\it not} to integrate over
$\widehat Q_0$.

\subsection{One-loop determinant}\la{ref:oneloopdet}

First we neglect all interactions, i.e.\ the term $R$
in~(\ref{expansioneff}), leaving the determinant without the zero mode
$\widehat Q_0$.

It is useful to generalize at this point the discussion from $SU(2)$
to $SU(N)$. For $SU(N)$, the $N-1$ independent eigenvalues are fixed by a product
of $N-1$ delta functions; their respective arguments are
\be
 \d\bigg(\ell_n-{1\over N}\Tr\overline{({\bf L}(A_0))^n}\bigg)\,,
~~~~~1\le n \le N-1\,.
\ee
The generalization of the $\epsilon$ field to $SU(N)$ follows
immediately: it is such that it couples to the Polyakov loop winding
$n$ times around the thermal circle. To fix the eigenvalues we need
$N-1$ windings $\Tr {\bf L}^n$. Hence, there are $N-1$ fields $\e_n$
and they generate the delta function constraints via the following
term in the exponent:
\be
i\sum_n\e_n \left(\ell_n-\Tr\overline{({\bf L}({A_0}))^n}\right)\,
\ee
with
\be
\Tr\overline{({\bf L}(A_0))^n}\equiv{1\over{V}}\int_V d\vec x~\Tr({\bf
  L}(A_0))^n\,.
\la{spaceav}
\ee
After expanding around the saddle point, the analogue of $\hat Q_0$ in
Eq.~(\ref{hatzeromom}) is labeled by the number of windings,
\be \widehat Q_0^{n}\equiv\Tr\bigg(\overline{\overline{Q}}_0({\bf
  L}^n({\bf q}))'\bigg)\equiv \sum_d\overline{\overline{Q}}_0^{\,d}~t^n_d({\bf
  q})\,.
   \la{hatzeromomn}
\ee
We introduced the matrix $t^n_d({\bf q})\equiv\Tr(\l_d ({\bf L}^n({\bf
  q}))')$ for later use. It connects the winding basis labeled by $n$
to the diagonal Cartan basis labeled by $d$.

By analogy to $SU(2)$, the $\ell_n$ can be written as
\be
\ell_n = {1\over N} \Tr \exp(2\pi in{\bf q})\,.
\ee
The matrix ${\bf q}$ is taken to be diagonal with its $N$ eigenvalues
$q_j$ obeying $\sum_j q_j=0$.

The diagonalization of $S''_{gauge}$ is
well known~\cite{Gross:1980br,Weiss:1980rj} and simply employs the
plane wave basis $Q_{\m}(p_0,\vec p)$. The color basis is the
well-known Cartan basis, spelled out in Sec.~(\ref{sec:twoloop})
for all classical groups. In this section we just need one salient
property of this basis.

It is divided into diagonal elements $H_d$ and off-diagonal elements
$E_\a$, eigenmatrices of the $H_d$,
\be
[H_d,E_\a]=\a_dE_\a.
\la{eigenmatrix}
\ee
The coefficients $\a_d$ are the components of an
$r$-dimensional vector, the root $\vec\a$, where $r$  is the rank of the group.

For the fundamental representation of $SU(N)$, this is as
follows. The fluctuation variables are labeled by
$d=1,2,\cdots, N-1$ corresponding to the $N-1$ diagonal
matrices. The off-diagonal fluctuation variables correspond to the
$N(N-1)$ off-diagonal matrices $\l^{ij}$, where the indices $i,j
=1,2,\cdots,N$ with $i \neq j$. The off-diagonal matrices are a direct
generalization of the off-diagonal Pauli matrices for $SU(2)$,
\be
  (\l^{ij})_{kl}={1\over{\sqrt{2}}}\d_{ik}\d_{jl}\,.
\la{offdiagsun}
\ee
As a consequence the Matsubara frequency in $D_0({\bf q})$,
Eq.~(\ref{gaugefix}), acting on the off-diagonal mode
$Q^{ij}_\m(p_0,\cdots)$ is shifted by $q_i-q_j\equiv q_{ij}$ but
remains unchanged if it acts on a diagonal mode $Q_\m^d$,
\ba
D_0({\bf q})Q^{ij}_\m(p_0,\cdots)&=&i(p_0+2\pi
q_{ij})Q^{ij}_\m(p_0,\cdots)\,, \nonumber\\
D_0({\bf q})Q^{d}_\m(p_0,\cdots)&=&ip_0Q^{d}_\m(p_0,\cdots)\,.
\la{covariantderiv}
\ea
Hence the shifted four-momenta are either $p^{ij}=(p_0+q_{ij},\vec p)$, or
$p^d=(p_0, \vec p)$, with $p_0=2\pi T n_0$ where $n_0$ is an integer. We
will use this notation throughout the paper. These rules generalize
to any other classical group (see Sec.~\ref{sec:twoloop}).


After these preliminaries we can easily compute the one loop
determinant; in dimensional regularization we obtain the well-knowm
result~\cite{Gross:1980br,Weiss:1980rj},
\be
\G_f=-p_{SB}+{2\pi^2\over 3}T^4\sum_{i\neq j}B_4(q_{ij}).
\la{uf}
\ee
The Bernoulli polynomial $B_4$ is given  in Appendix A.

\subsection{Interactions without the $\e$ fields}

These interactions are fully contained in the
interaction terms in the first line of~(\ref{expansioninteract}). They give
the usual free-energy diagrams $\G_f$ as in Fig.~\ref{fig:xi2loop} with
propagators and vertices determined by $\ell=\cos(2\pi q)$. Only
the zero momentum mode $\widehat Q_0$ is not integrated over.

The Feynman rules in the presence of the color diagonal background
${\bf q}$ are simple. They have been discussed in the previous section
and amount to replacing the momenta
as in  Eq.~(\ref{covariantderiv}) and below.

With these rules it is straightforward to obtain the contribution
$\G_f$ due to the free-energy diagrams in Fig.~\ref{fig:xi2loop} to
one and two loop order in Feynman gauge $\xi=1$,
\ba
\G_f=-p_{SB}+\sum_{a}\widehat B_4(q_a)+g^2\sum_{a,b,c}|{f^{a,b,c}}|^2\widehat B_2(q_b)\widehat B_2(q_c)\,.
\la{uf2}
\ea
The first two terms correspond to the one loop result~(\ref{uf}), the last term
is the two loop correction. The indices $a,b,c$ run through the
diagonal indices $d$ and the off-diagonal indices $ij$. So, $q_d=0$ and
$q_{ij}=q_i-q_j$. The Bernoulli polynomials are simple and defined in Appendix A.

\subsection{The insertion diagram due to the constraint}\la{subsec:insertion}

We now consider the interactions involving the fluctuations $\e_q$ in
the second line of Eq.~(\ref{expansioninteract}). They play an essential role
at two and more loops~\cite{KorthalsAltes:1993ca}.
They originate in terms
$S_{con}^{(3)}$, $(S_{con}^{(3)})^2$ etc. Among other contributions
the latter gives the term we wrote explicitly,
\be g^2 i\e_q \; \Tr \left(\overline{Q_0^2}\cdot {\bf L}''({\bf
  q})\right) \;
\left( Q^3\cdot S'''_{gauge}({\bf q}) \right)~.  \la{insertion0}
\ee
To avoid inessential complications we first discuss $SU(2)$.

The first factor in this expression is the Polyakov loop expanded to
second order in the fluctuation $Q_0$.  To two-loop order, this is the
only term that contributes at $\OO(V)$ to $\VV$. No other term does
because of the absence of infrared divergences.~\footnote{At three-loop order there are,
 however, linear infrared divergences.}

To perform the integration over $\e_q$, we replace $i\e_q$ by
${\partial\over{\partial \widehat Q_0}}$ acting on the last term
in~(\ref{expansioneff}). By inspection, the only contraction
of $\OO(V)$ is
\be
\left< \overline{Q_0^2}\cdot \Tr{\bf L}''({\bf
  q})\right>\,
\left<{\partial\over{\partial \widehat Q_0}} \; Q^3\cdot S'''_{gauge}({\bf
  q})\right> \la{insertion}\,.
\ee
The first contraction is the expectation value of the Polyakov loop
through one gluon exchange. The second contraction is the one point
function at zero momentum. The one point function is nonzero in
thermal physics because Lorentz invariance is reduced to
rotational invariance, so $Q_0$ is a scalar. It is shown in
Fig.~\ref{fig:insertion}.

\begin{figure}[htb]
\begin{center}
\includegraphics[width=6cm]{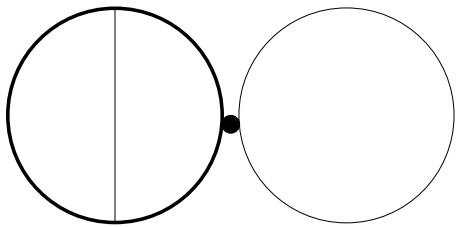}
\end{center}
\caption{The insertion diagram, Eq.~(\ref{insertion}). The two blobs
  correspond to the two contractions. The Polyakov loop is the fat
  circle, with the gluon going across. The dot is where $\widehat
  Q_0$ acts to create the one point function shown by the thin
  circle.}  \la{fig:insertion}
\end{figure}
This term would vanish if ${\bf L}''({\bf q})=0$, i.e.\ if the
constraint were linear. But it is not or else it would
not be gauge invariant. Indeed, in addition to the usual free-energy terms, this contribution renders the total result independent
of the gauge choice. This will become clear from the BRS analysis below.

In the $SU(2)$ case the spatial averages and the Polyakov loop average
simplify to
\ba
\< \Tr \overline{Q_0^2}\cdot {\bf L}''({\bf q})\> &=& (3-\xi)\widehat
B_1(q_{12}) \sin(\pi q_{12})\,,\nonumber\\
\widehat Q_0&=& \Tr (\overline{ \overline{Q}}_0 \sigma_3)\sin(\pi q_{12})\,,
\ea
where we used Eq.~(\ref{hatzeromom}). Also, Eq.~(\ref{insertion}) becomes
\be
\left< \Tr \overline{Q_0^2}\cdot{\bf L}''({\bf q})\right> \,
\left<{\partial\over{\partial \widehat Q_0}}\;
Q^3\cdot S'''_{gauge}({\bf q})\right> =
4(3-\xi)\widehat B_1(q_{12})\widehat B_3(q_{12})\,.
\ee
Note that all reference to the unitary nature of the loop in the
constraint has dropped out. The derivative of the loop, $\sin(\pi
q_{12})$ drops out of the insertion of the radiative correction for the
Polyakov loop into the one loop effective action. For groups larger
than $SU(2)$ this is true as well, though much less trivial (see Appendix~B).

In the $SU(N)$ case the contribution~(\ref{insertion}) becomes a sum
over $N-1$ winding contributions,
\be
\G_i=\sum_n\left< \overline{Q_0^2}\cdot \Tr{\bf L^n({\bf q})}''\right>\,
\left<{\partial\over{\partial \widehat Q^n_0}} \; Q^3\cdot S'''_{gauge}({\bf
  q})\right> \la{insertionsun}\,.
\ee

In Appendix~\ref{appb} we prove the crucial identity
\be
\left< \Tr (\overline{Q_0^2}\cdot({\bf L}^n)'')\right> =
\left< \Tr \overline{Q_0^2}\cdot{\bf L}''({\bf q})\right>_d\cdot
t^n_d({\bf q})\,.
\la{identitytwoloop}
\ee
This identity relates the one-loop expectation values of multiply
winding loops to that for single winding, using the matrix defined in
(\ref{hatzeromomn}).  In conjunction with~(\ref{hatzeromomn}) it eliminates
the summation over windings $n$ in~(\ref{insertionsun}) and reduces it
to a summation over diagonal indices $d$.
\be
\G^{(2)}_i =(3-\xi) \; g^2 \sum_{d,b,c}f^{d,b,-b}f^{d,c,-c}\; \widehat
B_1(q_b)\; \widehat B_3(q_c)\,.
\la{oneloopinsertfinal1}
\ee
We used for this result that the  VEV for single
winding is the $\OO(g^2)$ correction to the Polyakov loop $\Tr {\bf
  L}({\bf q})$ in the background ${\bf q}$,
  \be
  \left< \Tr \overline{Q_0^2}\cdot{\bf L}''({\bf q})\right>_d= (3-\xi)\sum_{d,ij}f^{d,ij,ji}\widehat{B}_1(q_{ij})\,.
\la{identitytwoloop1}
\ee
Note that this expectation value refers to the traced loop without the
normalization factor $1/N$. Further, that it is gauge choice dependent
and proportional to the Bernoulli function $\widehat
B_1(q_{ij})$. This function is linear, periodic mod 1, antisymmetric
and vanishes at $q_{ij}=1/2$. Hence it is a sawtooth function, with
nonzero values at $q_{ij}=0$, 1.
As usual, $d$ refers to the diagonal index while $b$ and $c$ run through
off-diagonal indices only.  In Sec.~\ref{sim} we simplify the summation over
the index $d$.

As $\widehat B_3$ vanishes linearly for small argument it follows that
$\G_i^{(2)}$ vanishes, to this order, in the limit of zero
background.

\subsection{BRS identities and the gauge independence of the effective potential}{\la{subsec:brs}}

We need to understand {\it why} the $\xi$ dependence in the diagram for the
free-energy
contribution $\G_f$ cancels against that in the insertion
diagram $\G_i$.
There is a simple way to see this, using  the BRS identities in the presence of a thermal background~\cite{KorthalsAltes:1993ca}.

\begin{figure}[htb]
\begin{center}
\includegraphics[width=10cm]{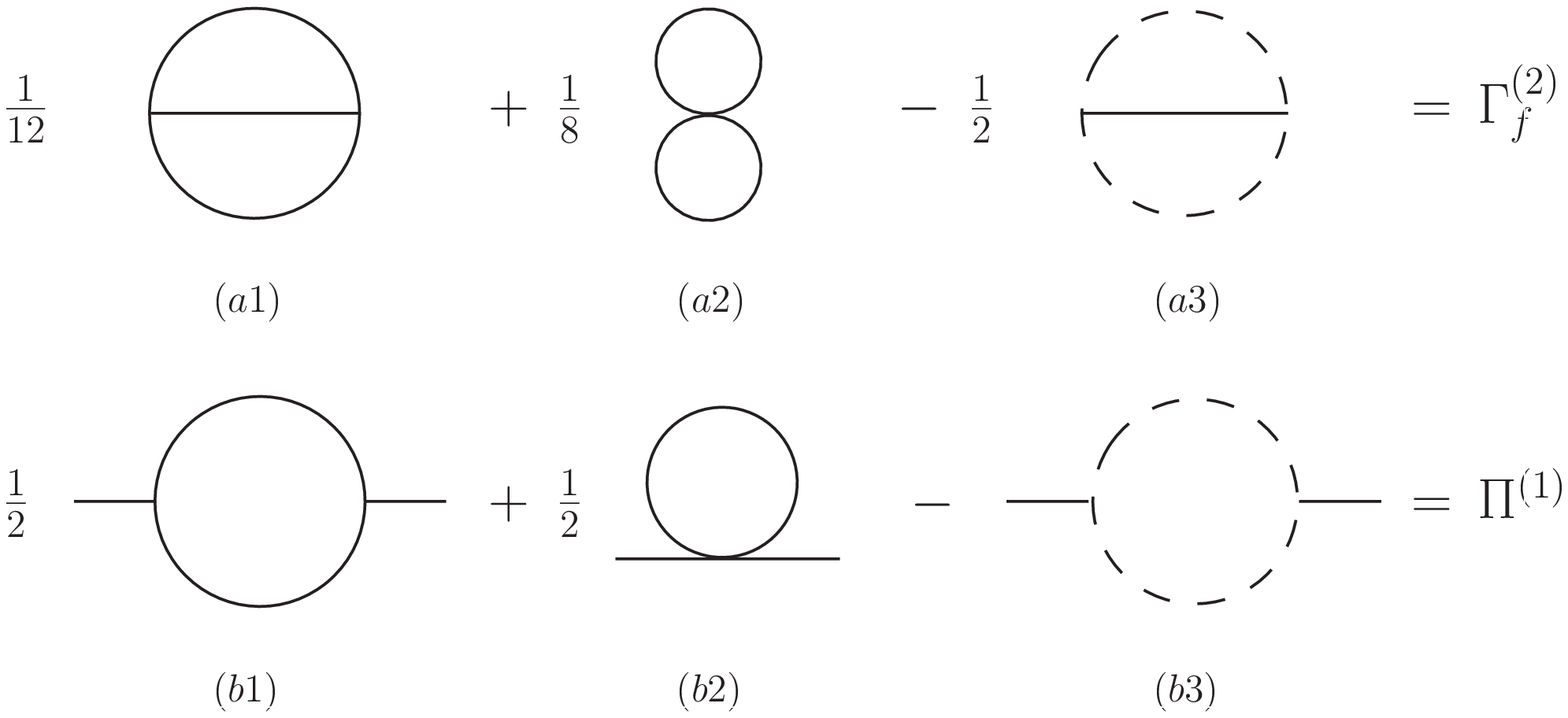}
\end{center}
\caption{The two-loop free-energy contributions $\G^{(2)}_f$ to the
  effective potential are shown in (a1), (a2) and (a3). The one-loop
  self-energy is shown in (b1) - (b3).}  \la{fig:xi2loop}
\end{figure}
All we have to do is to take the free-energy contribution for an
arbitrary value of $\xi$ and to note that the $\xi$ dependence is due
exclusively to the gluon propagators.
It does not appear anywhere else in the free-energy diagrams.

Varying the gluon propagators in the three diagrams (a1), (a2) and
(a3) shown in Fig.~(\ref{fig:xi2loop}) multiplies each by 3, 2 and 1,
respectively. Combining these factors with the combinatorial factors
in the figure turns the result into the one loop gluon self-energy
$\Pi^{(1)a,b}_{\m,\n}$ [see (b1) - (b3) in Fig.~\ref{fig:xi2loop}]
folded into the gauge part of the propagator,~\footnote{This identity
  is a special case of an identity valid for any field theory.}
\be
{\partial \G^{(2)}_f\over{\partial \xi}}=\sumint_p
\Pi^{(1)a,b}_{\m,\n} {p^a_\m p^b_\n\over{(p^a)^4}} \la{gaugevaruf}\,.
\ee
We use the shorthand notation
\be T\sum_{n_0}\int {d^{d-1}\vec
  p\over{(2\pi)^{d-1}}} \equiv \sumint_p\,. \la{shorthand}
\ee
If $a=b=d$, with $d$ a diagonal index, the BRS identity tells us, as
in the case without background, that the one-loop self-energy is
transverse,
\be
\Pi^{(1)d,d}_{\m,\n} \, p^d_\m p^d_\n=0\,.
\la{gaugevartwotozero}
\ee
In case $a=(ij), b=(ji)$ is off-diagonal the BRS identity relates the
two-point function to the one-point function $\G_d$,
\be
\G^{(1)}_d=\left<{\partial{S_{int}}\over{\partial{Q_0^d}}}\right>~.
\ee
Only the scalar, color diagonal one-point function is nonvanishing in
thermal field theory. We obtain
\ba
\Pi^{(1)ij,ji}_{\m,\n} \; p^{ij}_\m p^{ji}_\n &=&
f^{d,ij,ji} p^{ij}_0\; \G^{(1)}_d({\bf q})\,,\nonumber\\
\G^{(1)}_d(q)&=& \sum_{k,l}f^{d,kl,lk}\widehat B_3(q_{kl})\,.
\la{brstwotoone}
\ea

Unless explicitly shown there is no summation over color indices in (\ref{gaugevartwotozero}) and (\ref{brstwotoone}).
The second equality relates the one point function to the background
field derivative of the free energy.
The final result for the gauge variation follows from
Eq.~(\ref{gaugevaruf})
and the BRS identities~(\ref{gaugevartwotozero}) and~(\ref{brstwotoone}),
\be
{\partial \G^{(2)}_f\over{\partial \xi}}=g^2\sum_{ijkl}\sumint_p
{p_0^{ij}\over{(p^{ij})}^4}\sum_df^{d,ij,ji}f^{d,kl,lk}\widehat B_3(q_{kl})\,.
\ee
Inspection of the first factor in this expression shows (see Appendix~\ref{appa}
on Bernoulli functions) that
\be
\sumint_p {p_0^{ij}\over{(p^{ij})}^4}=\widehat B_1(q_{ij})\,.
\ee
Hence, the gauge dependence cancels precisely with the gauge
variation of the insertion diagram~(\ref{oneloopinsertfinal1}).

\section{TWO-LOOP CORRECTION: EXPLICIT RESULT}\la{sec:twoloop}

Now that we have seen the cancellation of the gauge artifacts in the
two contributions to the effective action, we evaluate them for various
groups.  Because the free-energy contribution $\G_f$, Eq.~(\ref{uf}),
is so simple in $\xi=1$ gauge, we calculate
Eq.~(\ref{oneloopinsertfinal1}) for $\xi=1$ as well.  It shows that
the insertion diagram not only guarantees gauge parameter
independence, but also a surprisingly simple outcome for the effective
action.

We should mention also that although the specific forms of Eqs.~(\ref{uf})
and~(\ref{oneloopinsertfinal1}) are based on our discussion of
$SU(N)$, they in fact apply to all the groups that we are going to
consider. For later use, we rewrite the effective potential up to two-loop
order with $\xi=1$ as
\ba
\la{1loop}
\G^{(1)}&=& -\frac{\pi^2 T^4 d(A)}{45}+\sum_{a}\widehat B_4(q_a)\, ,\\
\la{newuf}
\G^{(2)}_f&=& g^2\sum_{a,b,c}|{f^{a,b,c}}|^2\widehat B_2(q_b)\widehat B_2(q_c)\, ,\\
\la{newoneloopinsertfinal1}
\G^{(2)}_i&=&2 g^2\sum_{d,b,c}f^{d,b,-b}f^{d,c,-c}\widehat B_1(q_{b})\widehat B_3(q_{c})\,.
\ea
Here, the one-loop effective action depends on the dimension of the
adjoint representation of the group which is denoted as $d(A)$. It
equals $N^2-1$ for $SU(N)$, $2N^2-N$ for $SO(2N)$, $14$ for $G(2)$,
while for both $Sp(N)$ and $SO(2N+1)$ we have $d(A)=2N^2+N$. The index
$a$ runs through the off-diagonal indices. In $\G^{(2)}_f$, the
indices $a$,$b$ and $c$ run over both diagonal and off-diagonal
indices. In $\G^{(2)}_i$, each structure constant contains the
diagonal indices $d$, while $b$ and $c$ denote off-diagonal
indices. If $b$ is a typical off-diagonal index, the index $-b$ is
defined as follows: let $E^{b}$ being some off-diagonal generator,
then $E^{-b}\equiv (E^{b})^{\dagger}$. The definition of these indices
will become more clear in the following.

Our calculation will show that the two-loop effective potential is
simply a multiplicative and background {\it independent}
renormalization of the one-loop result. In terms of the
quadratic Casimir invariant $C_2(A)$ in the adjoint representation,
\be
\frac{\G^{(2)}}{\G^{(1)}}=-{5 g^2C_2(A)\over{16\pi^2}} ~,
\la{twoloopresult}
\ee
where the Casimir invariant is given by
\be
C_2(A)\; \d_{ce}=f^{a,b,c}f^{a,b,e}~.
\la{casimir}
\ee
From this definition it follows in particular that for two diagonal
indices $c=d$ and $e=d^\prime$,
\be
C_2(A)\; d(r) = f^{a,b,d}f^{a,b,d}~,
\la{casimirroots}
\ee
where $d(r)$ is the rank of the group, i.e.\ the dimension of the
Cartan space.  We have that $C_2(A)= N-1$ for $SO(2N)$, $N-{1\over 2}$
for $SO(2N+1)$, $N+1$ for $Sp(2N)$, $2$ for $G(2)$, and finally
$C_2(A)=N$ for the $SU(N)$ groups. For the $SU(N)$ groups the
result~(\ref{twoloopresult}) was in fact known since long for straight
paths\footnote{However, in general the minimum of the potential does
  not exactly follow a straight path as a function of
  temperature~\cite{Dumitru:2012fw}.} from the origin, ${\bf q}=0$, to
the degenerate $Z(N)$ minima~\cite{Bhattacharya:1992qb}. These paths
run along the edges of the $SU(N)$ Weyl chamber and a combinatorial
proof of Eq.~(\ref{twoloopresult}) exists~\cite{Giovannangeli:2002uv}.
We do not (yet) know how the combinatorics works out inside the Weyl
chamber.

Below we gather the tools to produce explicit expressions for the two
loop insertion $\G^{(2)}_i$ and the free energy $\G^{(2)}_f$. Due to
the increasing number of independent variables of the background field
and the complication of the indices of the structure constants, we
developed a \textsc{Mathematica} program~\cite{M6-8} for all classical groups
to evaluate explicitly the above two contributions, Eq.~(\ref{newuf})
and Eq.~(\ref{newoneloopinsertfinal1}). However, we have not succeeded
in finding a general proof of Eq.~(\ref{twoloopresult}) which does not
require explicit evaluation by brute force.

\subsection{Generalities on the classical Lie algebras}
We start with the commutation relations in the Cartan basis for any semi-simple Lie algebra,
\ba
~[\vec H,E_{\a}]&=&\vec\a ~E_{\a} \\
~[E_{\a},E_{-\a}]&=&\vec\a \cdot \vec H\\
~[E_{\a},E_{\b}]&=&f^{\a,\b,-\a-\b}E_{\a+\b}\, ,
~\mbox{if}~ \a+\b~\mbox{is a root; if not, it vanishes.}
 \la{eq:appcommrel}
\ea
We define the structure constants from the generators in the
fundamental representation of the group, with the generators normalized
as
\be
\Tr (E_\a E_{-\a})=\Tr (H_d^2)=1/2~.
\ee
The components of $\vec H$ are the orthonormal matrices spanning the
Cartan subalgebra and they are the diagonal generators in the Cartan
basis. The orthonormal $E_\a$, labelled by the roots $\a$, are vectors
in Cartan space. They are the off-diagonal generators.

The roots themselves are labeled by an off-diagonal index. For a
typical off-diagonal index, say $a$, we have
\be
~[H^d,E_a]=f^{d,a,-a}E_a.
\ee
Hence, the $d$th component of a root (labeled by an off-diagonal
index $a$) is the structure constant $f^{d,a,-a}$. Besides the
structure constants involving a diagonal component, we have another
kind of structure constants $f^{\a,\b,-\a-\b}$ which connect
off-diagonal generators. With our normalization, the absolute values
of $f^{\a,\b,-\a-\b}$ are all equal to ${1\over{\sqrt{2}}}$ for
$SU(N)$. For all the other classical groups they are ${1\over 2}$. For
the exceptional group $G(2)$, they are given in Sec.~\ref{g2}. In the
following, we shall discuss the commutation relations of the
generators and the corresponding structure constants for each group
separately.

\subsection{Calculation for $SU(N)$}
\la{sun}

Starting from Eqs.~(\ref{newuf}) and~(\ref{newoneloopinsertfinal1}),
we are able to calculate the two-loop perturbative correction to the
effective potential. First of all, we need to know the structure
constants. They can be obtained from the commutation relations of the
generators in Cartan basis.

For $SU(N)$, there are $N(N-1)$ off-diagonal generators
$E^{ij}\equiv\l^{ij}$ with $i,j=1,\cdots,N$ and $i \neq j$. The
explicit forms are given in Eq.~(\ref{offdiagsun}). In addition, we
have $N-1$ traceless diagonal generators $H^d\equiv\l^{d}$ with
$d=1,\cdots,N-1$,
\be
\l^{d}=\frac{1}{\sqrt{2d(d+1)}}\,
{\rm diag}(1,1,\cdots\, ,-d,0,0,\cdots\,,0)\, .
\ee
The commutators between diagonal generators are obviously zero. The
nonvanishing commutators we need are\footnote{For $SU(N)$, if a
  typical off-diagonal index $b$ is denoted by $b=ij$, then we have
  $-b=ji$.}
\ba
\la{com1}
[H^{d},E^{ij}]&\equiv&f^{d,ij,lk}E^{kl}=(\l^d_{ii}-\l^d_{jj})
E^{ij}\,,\\{}
[E^{ij},E^{kl}]&\equiv&f^{ij,kl,ts}E^{st}= \frac{1}{\sqrt{2}}(\d_{jk}E^{il}-\d_{il}E^{kj})\,.
\la{com}
\ea
Here, $\l^d_{ii}$ is the $i$th diagonal component of $\l^d$. From
Eq.~(\ref{com1}) we find that the roots ${\vec \a}^{\,ij}=({\vec
  \l}_{ii}-{\vec \l}_{jj})$. As mentioned before, the $d$th
component of ${\vec \a}^{\,ij}$ is the structure constant
$f^{d,ij,ji}$.

We can define a diagonal matrix,
\be
\la{dl}
\L^{ij}\equiv \frac{{\vec \l}\cdot{\vec \a}^{\,ij}}{({\vec \a}^{\,ij})^2}\, ,
\ee
and it is easily to find the following commutator:
\ba
[\L^{ij},E^{ij}]&=& E^{ij}\, .
\ea
Using the explicit form of $E^{ij}$, we get $\L^{ij}=\frac{1}{2} {\rm
  diag}(0,0,\cdots,1,0,\cdots,0,-1,0,\cdots,0)$, i.e.\ the $i$th
component is $1$, the $j$th component is $-1$ and all others are zero.

Taking the square of Eq.~(\ref{dl}) and then the trace on both
sides, the roots satisfy
\be
\la{roots}
({\vec \a}^{\,ij})^2=1\, .
\ee
In other words, the roots can be written in terms of an orthonormal
basis $\{{\vec e}_i\}$ spanning an $N$-dimensional space,
\be
{\vec \a}^{\,ij}=\frac{1}{\sqrt{2}}({\vec e}_i-{\vec e}_j)\, .
\ee
Using Eqs.~(\ref{casimirroots}) and~(\ref{roots}), we have $C_2(A)=N$
for $SU(N)$. Notice that there are $N^2-N$ off-diagonal indices and
that the rank of $SU(N)$ is $N-1$.

Using the total antisymmetry of the structure constants, all the
nonvanishing structure constants without diagonal index can be
read off from Eq.~(\ref{com}). It is obvious that the absolute values
of these structure constants are $1/\sqrt{2}$.

Since the explicit form of the generators is known, the calculations
of the structure constants is straightforward but rather tedious as
$N$ becomes large. In fact, we can rewrite Eqs.~(\ref{com1})
and~(\ref{com}) to obtain the following expressions for the structure
constants
\ba
f^{d,ij,kl}&=&2 {\rm Tr}(E^{kl}\cdot[H^{d},E^{ij}])\,,\nonumber\\{}
f^{ij,kl,st}&=&2 {\rm Tr}(E^{st}\cdot[E^{ij},E^{kl}])\,.
\la{deff}
\ea
These expressions permit a straightforward computation of the
structure constants using \textsc{Mathematica}.

When using Eqs.~(\ref{1loop}) to~(\ref{newoneloopinsertfinal1}) to
compute the effective potential for $SU(N)$, one needs to observe that
for a diagonal index $q_a=0$.  However, for an off-diagonal index $ij$
one has $q_a= q_i - q_j$. We have $N-1$ independent $q_i$, for
$i=1,2,\cdots,N-1$. Thus, the background field can be parametrized as
$Q={\rm diag}(q_1,q_2,\cdots\, ,q_N)$ with the single constraint
$q_N=-q_1-q_2-\dots\,-q_{N-1}$. The above discussion can be understood
by using the following commutator:
\be
[Q,E^{ij}] = (q_i-q_j) E^{ij}\, .
\ee
It is obvious that $Q$ commutes with $H^d$ which leads to $q_d=0$ as
stated.

In general, there is no restriction to the possible values of $q_i$
with $i=1,2,\cdots, N-1$. Therefore, modulo functions appear in the
Bernoulli polynomials which makes the calculation more
involved. However, without loss of generality, we can perform the
calculation with a set of $q^\prime_i$ such that the absolute values
of the arguments of the Bernoulli polynomials are less than 1, i.e.,
$-1<q^\prime_i-q^\prime_j<1$. It is easy to show that it is always
possible to find such a set of $q^{\prime}_i$. For example, when
considering $SU(3)$, we have $Q={\rm diag}(q_1,q_2,q_3)$ with
$q_3=-q_1-q_2$. If we define $q_i^\prime = q_i-n_i$ where $n_i$ is an
integer, we can achieve $0 \le q_i^\prime < 1$ by appropriate choice
of $n_i$. Since the Bernoulli polynomials are periodic modulo 1, one
can use $q^\prime_i$ instead of $q_i$ to calculate the effective
potential and the result is the same. The advantage of using
$q^\prime_i$ is to avoid these modulo functions (see
Appendix~\ref{appa}) which can not be handled easily by
\textsc{Mathematica}. We mention that in terms of $q^\prime_i$ the background
field is not necessarily traceless. In fact, we have $q^\prime_3 =
-q^\prime_1-q^\prime_2+n$ with $n=0,~1,~2$. With the set of
$q^\prime_i$, the Bernoulli polynomials are given by
Eq.~(\ref{bernoulli4}) with sign functions only. By permutation of the
matrix elements of the background field, we can assume that
$q^\prime_1\ge q^\prime_2\ge\cdots\ge q^\prime_N$. With this
assumption, the sign of $q^\prime_i-q^\prime_j$ becomes
definitive which can further simplify the calculation by ignoring the sign
functions in the Bernoulli polynomials. The details can be found in Appendix~\ref{appa}.

Based on Eq.~(\ref{deff}) and the above discussion, we have been able
to compute the two-loop perturbative correction to the effective
potential for $SU(N)$ for any given but arbitrary $N$; we have performed
this calculation explicitly up to $N=5$ with \textsc{Mathematica}~\cite{M6-8}
and verified the relation~(\ref{twoloopresult}). For example, for
$SU(2)$ we find
\ba \Gamma^{(2)}_f&=&\frac{g^2 T^4}{24}[1 + 2
  q^\prime_1(q^\prime_1-1) + 2 q^\prime_2(1+q^\prime_2-2
  q^\prime_1)][1 + 6 q^\prime_1(q^\prime_1-1)\nonumber\\ && + 6
  q^\prime_2(1+q^\prime_2-2 q^\prime_1)]\, , \nonumber
\\ \Gamma^{(2)}_i&=& \frac{g^2 T^4}{3} (1 + 2 q^\prime_2 - 2
q^\prime_1)^2 (q^\prime_1 - q^\prime_2) (1 + q^\prime_2 -
q^\prime_1)\, .  \la{su22loop}
\ea
The effective potential at one-loop order is given by
\be
\Gamma^{(1)}=-\frac{\pi^2 T^4 }{15}+\frac{4 T^4 \pi^2}{3} (q^\prime_2
- q^\prime_1)^2 (1 + q^\prime_2 - q^\prime_1)^2\, ,  \la{su21loop}
\ee
so that
\be
\frac{\Gamma^{(2)}_f+\Gamma^{(2)}_i}{\Gamma^{(1)}}=-\frac{5g^2}{8
  \pi^2}\, ,\,\,\,\,\, {\rm for}\,\, N=2\,.
\ee
Note that we write these equations in terms of $q^\prime_i$ to
remind the readers that these variables should satisfy $0 \le
q_i^\prime < 1$ and $q^\prime_1\ge q^\prime_2\ge\cdots\ge
q^\prime_N$. For example, if the background field is given as $Q={\rm
  diag}(-\frac{5}{3},\frac{5}{3})$, to get the correct effective
potential from Eqs.~(\ref{su22loop}) and~(\ref{su21loop}) one should
set $q^\prime_1=\frac{2}{3}$ and $q^\prime_2=\frac{1}{3}$.  This
procedure can be easily generalized to higher $N$.

\subsection{Calculation for $SO(2N)$ and $SO(2N+1)$}

For these groups we use a variant of the notation from Georgi's
book~\cite{georgi}. The generators $M^{ab}$ in the fundamental
representation have matrix elements,
\be
(M^{ab})_{xy}=-\frac{i}{2}(\d_{ax}\d_{by}-\d_{ay}\d_{bx})\, .
\ee
Obviously there is antisymmetry under exchange of the labels $a$ and
$b$, i.e.\ $M^{ab}=-M^{ba}$.

Furthermore, we can define the off-diagonal generators in the Cartan
basis. For both groups, there are $N(2N-2)$ off-diagonal generators
$E^{\eta i. \eta^\prime j}$ with $i,j=1,\cdots,N$ and $i>j$. Here, we
define the indices $i$ with an associated sign $\eta$. Similarly, $j$
is defined with $\eta^\prime$. The signs $\eta$ or $\eta^\prime$ are
independently $\pm 1$. The explicit form of the generators is
\be E^{\eta i. \eta^\prime j}=\frac{1}{2}(M^{2i-1,2j-1}+i \eta
M^{2i,2j-1}+i \eta^\prime M^{2i-1,2j}-\eta \eta^\prime M^{2i,2j})\,.
\ee
For $SO(2N+1)$ there are $2N$ additional off-diagonal generators
\be
E^{\eta i}=\frac{1}{\sqrt{2}}(M^{2i-1,2N+1}+i \eta M^{2i,2N+1})\,.
\ee
For either of the groups the $N$-dimensional Cartan subalgebra is spanned by
mutually commuting and orthogonal generators $H^d$, with
\be
H^d=M^{2d-1,2d}\, ,\, {\rm with} \,\,\,d=1,2,\cdots\,N.
\la{eq:diagcartan}
\ee

So far, we have defined all the generators in the Cartan basis; the
structure constants can be obtained from the commutation
relation\footnote{For $SO(2N)$ and $SO(2N+1)$, if the typical
  off-diagonal index $b$ is denoted as $b=\eta i.\eta^\prime j$ then
  $-b=-\eta i.-\eta^\prime j$; if $b=\eta i$, then $-b=-\eta i$. In
  Eq.~(\ref{4com}), with our notation, $E^{\rho k.\eta i}$ should be
  understood as $-E^{\eta i.\rho k}$ if $i>k$. Similarly for
  $E^{\eta^\prime j.\rho^\prime l}$.}
\ba \la{1com} [H^d,E^{\eta j}]&\equiv& f^{d,\eta j,-\eta^\prime k}
E^{\eta^\prime k}=\frac{\eta}{2}\d_{dj} E^{\eta j}\\{}\la{2com}
[H^d,E^{\eta j.\eta^\prime k}]&\equiv& f^{d,\eta j.\eta^\prime k
  ,-\rho l.-\rho^\prime m} E^{\rho l.\rho^\prime
  m}=\frac{1}{2}(\eta\d_{dj}+\eta^\prime\d_{dk}) E^{\eta j.\eta^\prime
  k}\\{}\la{3com} [E^{\eta i.\eta^\prime j},E^{\rho k}]&\equiv&
f^{\eta i.\eta^\prime j, \rho k, -\sigma l} E^{\sigma
  l}=\frac{i}{4}\bigg(\d_{ki}(1-\rho \eta)E^{\eta^\prime
  j}-\d_{kj}(1-\rho \eta^\prime)E^{\eta i}\bigg)\\{} [E^{\eta
    i.\eta^\prime j},E^{\rho k.\rho^\prime l}]&\equiv& f^{\eta
  i.\eta^\prime j,\rho k.\rho^\prime l,-\sigma t.-\sigma^\prime n}
E^{\sigma t.\sigma^\prime n}=\frac{i}{4}\bigg(\d_{ki}(1-\rho
\eta)E^{\eta^\prime j.\rho^\prime l} -\nonumber \\
&&\hspace{-1cm} \d_{kj} (1-\rho
\eta^\prime)E^{\eta i.\rho^\prime l}-\d_{lj}(1-\rho^\prime
\eta^\prime)E^{\rho k.\eta i}+\d_{il}(1-\eta \rho^\prime)E^{\rho
  k.\eta^\prime j}\bigg)\,. \la{4com} \ea
From Eqs.~(\ref{1com}) and (\ref{2com}), the roots can be expressed as
\ba
{\vec \a}^{\,\eta i}&=&\frac{\eta}{2}{\vec e}_i\, , \nonumber \\
{\vec \a}^{\,\eta i.\eta^\prime j}&=&\frac{1}{2}(\eta{\vec e}_i+\eta^\prime{\vec e}_j)\, .
\ea
There are $N(2N-2)$ off-diagonal generators associated with the long
roots and $2N$ off-diagonal generators associated with the short
roots. For both $SO(2N)$ and $SO(2N+1)$, $d(r)=N$. Using
Eq.~(\ref{casimirroots}) we can easily get $C_2(A)=N-\frac{1}{2}$ for
$SO(2N+1)$ and $C_2(A)=N-1$ for $SO(2N)$.

Like for $SU(N)$, in order to perform the calculation with \textsc{Mathematica}
we express the structure functions as
\ba
f^{d,\eta j,\eta^\prime k}&=&2 {\rm Tr}(E^{\eta^\prime k}\cdot [H^d,E^{\eta j}])\nonumber\\{}
f^{d,\eta j.\eta^\prime k,\rho l.\rho^\prime m}&=&2 {\rm Tr}(E^{\rho l.\rho^\prime m}\cdot[H^d,E^{\eta j.\eta^\prime k}])\nonumber\\{}
f^{\eta i.\eta^\prime j,\rho k,\sigma l}&=&2{\rm Tr}( E^{\sigma l}\cdot[E^{\eta i.\eta^\prime j},E^{\rho k}])\nonumber\\{} f^{\eta i.\eta^\prime j,\eta k.\eta^\prime l,\sigma t.\sigma^\prime n}&=&2{\rm Tr}(E^{\sigma t.\sigma^\prime n}\cdot[E^{\eta i.\eta^\prime j},E^{\rho k.\rho^\prime l}])
\la{deff2}
\ea
If an index $a$ is an off-diagonal index, we
have two different cases: if $a=\eta i$, then $q_a=\eta q_i$; if
$a=\eta i.\eta^\prime j$, then $q_a=\eta q_i+\eta^\prime q_j$. Here,
we have $N$ independent $q_i$ for $i=1,2,\cdots,N$. Thus, the
background field can be parametrized as $Q=\sum_{d=1}^N 2 q_d
H^d$. The above discussion can be understood by using the two
commutators from Eqs.~(\ref{1com}) and~(\ref{2com}).

In order to perform the computation with our \textsc{Mathematica} program we
again require an appropriate choice of $q_i$. For $SO(2N+1)$ and
$SO(2N)$, we can always start the calculation by using a set of $q_i$
which satisfy $-\frac{1}{2}<q_i\le\frac{1}{2}$. As a result, the
arguments of the Bernoulli polynomials are restricted to the interval
$-1\le x\le 1$, and we can use Eq.~(\ref{bernoulli4}) which does not
involve modulo functions. Without loss of generality, we can also
assume $q_1\ge q_2\ge\cdots\ge q_N\ge 0$ by a suitable permutation of
the matrix elements of $Q$. At this point we are able to use our
program to compute the effective potential for these two groups. For
example, with $N=2$, the results for $SO(5)$ are
\ba \Gamma^{(2)}_f&=&\frac{g^2 T^4}{8}
\biggl[\frac{5}{6}+(3q_1^2-3q_1) (2 - 3 q_1 +3 q_1^2) - (8
  q_1^2-8q_1+2) q_2 + (11 \nonumber \\&& -36 q_1 + 36 q_1^2) q_2^2 - 6
  q_2^3 + 9 q_2^4\biggr]\, , \nonumber \\ \Gamma^{(2)}_i&=& \frac{g^2
  T^4}{4}\biggl[3 (1 - 2 q_1)^2 (1 - q_1) q_1 + 4
  (q_1^2-q_1+\frac{1}{4}) q_2 + (48 q_1-13\nonumber \\ &&- 48 q_1^2)
  q_2^2 + 8 q_2^3 - 12 q_2^4\biggr]\, .  \la{s052loop}
\ea
The effective potential at one-loop order reads
\be \Gamma^{(1)}=-\frac{2 \pi^2 T^4 }{9}+\frac{4 T^4 \pi^2}{3}
\biggl[3 (q_1-1)^2 q_1^2 - 2 q_2^3 + 3 q_2^4 + 3 (q_2 - 2 q_1 q_2)^2
  \biggr]\, .  \la{so51loop}
\ee
It is straightforward to show that
\be
\frac{\Gamma^{(2)}_f+\Gamma^{(2)}_i}{\Gamma^{(1)}}=-\frac{15g^2}{32
  \pi^2}\, ,\,\,\,\,\, {\rm for}\,\, N=2\,.
\ee
We can easily get analogous results for $SO(2N)$ simply by ignoring
the off-diagonal generator $E^{\eta i}$.  We have verified
Eq.~(\ref{twoloopresult}) for $SO(2N+1)$ and $SO(2N)$ up to $N=5$.

\subsection{Calculation for $Sp(2N)$}

In this section we discuss the symplectic groups $Sp(2N)$. They are
the pseudoreal part of $SU(2N)$ constructed by defining the charge
conjugation matrix,
\be
  I_{2N}=i\s_2\otimes {\bf 1}_N ~, \la{eq:ccmatrix}
\ee
and requiring the special unitary matrix $U$ to obey
\be I_{2N}\; U\;
I_{2N}^\dagger=U^* ~,  \la{eq:symplectic}
\ee
where $\s_i$ are the Pauli matrices with $i=1,2,3$, and ${\bf 1}_N$ is
the $N$-dimensional unit matrix.

Writing $U={\rm exp}(i {\cal G})$, the symplectic generator is of the form
\ba
{\cal G}=\left (\begin{array}{cc} A&B\\
B^*&-A^*
\end{array}\right)
\label{eq:symgen}\, .
\ea
Here, $A$ is a Hermitian matrix with $A=A^\dagger$, and $B=B^t$ is
complex. For $N=1$, this form indeed reduces to the generator of
$SU(2)$. The Hermitian matrix A is not traceless, but ${\cal G}$
is. We therefore have $N^2$ real degrees of freedom from $A$, and
$N(N+1)$ degrees of freedom from the symmetric complex matrix $B$. In
total, we have $N(2N+1)$. The Cartan space is $N$ dimensional.

The diagonal generators of $Sp(2N)$ is
\ba
H^d={1\over{\sqrt{2}}}\; \s_3\otimes \l^d\,,~d=1,\cdots,N ~.
\ea
Here, the $N-1$ matrices $\l^d$ are the same as for $SU(N)$, and
we need the additional $\l^N={1\over{\sqrt{2N}}}{\bf 1}_N$.

The corresponding off-diagonal generators $E^{ij}$ are
\ba
E^{ij}={1\over{\sqrt{2}}}\left (\begin{array}{cc} \l^{ij}&0\\
0&-\l^{ji}
\end{array}\right)
\label{eq:symgendiag},~i,j=1,\cdots,N\,,\quad {\rm and}\quad i\neq j\,.
\ea
The $E^{ij}$ produce the roots of $SU(N)$, up to a factor of
$\frac{1}{\sqrt{2}}$,
\be
[H^{d},E^{ij}]=\frac{1}{\sqrt{2}}(\l^d_{ii}-\l^d_{jj})E^{ij}\,.
\ee
In addition, we have additional $N(N+1)$ off-diagonal generators of the
complex symmetric matrix $B$ which are denoted $E^{\eta ij}$; the
first index $\eta$ is a sign index. They are defined by
\ba
\la{Eije}
E^{\eta ij} &=&
\left[
  {1\over{\sqrt{2}}}+\d_{ij}(\frac{1}{2}-\frac{1}{\sqrt{2}})\right]
\; \s^{\eta}\otimes
(\l^{ij}+\l^{ji})\,, ~i,j=1,\cdots,N,\quad {\rm and}\quad i\ge j\,,
\ea
where $\s^{\eta}={1\over 2}(\s_1+ i \eta \s_2)$. Here, the index $i$
can be equal to $j$ which defines $2N$ long roots.
The generators $E^{\eta ij}$ produce a new type of roots. For $i>j$, we have
\be
[H^{d},E^{\eta ij}]=\frac{\eta}{\sqrt{2}}(\l^d_{ii}+\l^d_{jj})E^{\eta ij}\,.
\ee
For $i=j$, we have
\be
[H^{d},E^{\eta ii}]=\eta\sqrt{2}\l^d_{ii}E^{\eta ii}\,.
\ee

Like for $SU(N)$, we can write the roots for $Sp(2N)$
in terms of the orthonormal basis $\{{\vec e}_i\}$
introduced in Sec.~(\ref{sun}),
\ba
\vec\a^{\,\eta ij}&=& {\eta\over 2}( \vec e_i +\vec e_j)\,,~1\le j <
i\le N\, ,\nonumber\\
\vec\a^{\,ij}&=&{1\over 2}(\vec e_i-\vec e_j)\, ,~1\le i\le N\, ,~1\le
j\le N\, ~{\rm and}\,~ i\neq j\,,\nonumber\\
\vec\a^{\,\eta i}&=&\eta \vec e_i,~1\le i\le N\,.
\la{eq:rootsspninebasis}
\ea
Here, the first roots are associated with the generators $E^{\eta ij}$
when $i>j$ and the second roots are associated with the generators
$E^{ij}$. These two kinds of roots have length $\frac{1}{\sqrt{2}}$
and they are the short roots. There are $2N(N-1)$ of those. The roots
$\vec\a^{\,\eta i}$ (which can be also written as $\vec\a^{\,\eta
  ii}$) come from $E^{\eta ii}$. The $2N$ roots $\vec\a^{\,\eta i}$
have length $1$ and are the long roots.

Inversion of the roots is called duality and delivers the roots of
$SO(2N+1)$.  The roots of $Sp(2N)$ are those of $SO(2N+1)$, with the
long roots being the short ones, and the short roots being the long
ones. With our normalization of the generators there is an overall
factor of $1/2$.

From this discussion we can easily get $C_2(A)=N+1$ for $Sp(2N)$. For
$N=1$, we find $C_2(A)=2$ which is the same as for $SU(2)$, as expected.

On the other hand, the commutation relations between the off-diagonal
generators are
\ba
~[E^{ij},E^{kl}]&=&\frac{1}{2}(\d_{jk}E^{il}-\d_{il}E^{kj})\nonumber\,,\\
~[E^{+ij},E^{-kl}]&=&{1\over 2}(\d_{jk}E^{il}+\d_{il}E^{jk}+\d_{jl}E^{ik}+\d_{ik}E^{jl})\nonumber\,,\\
~[E^{\eta ij},E^{\eta kl}]&=&0\,.
\ea
The first line is the commutation relation of $SU(N)$, with the
structure constant $1/2$ instead of $1/\sqrt{2}$. The other lines all
reflect the symmetry in the indices of the $E^{\eta kl}$.

For $Sp(2N)$, we have four different types of structure constants due
to the nonvanishing commutators, namely, $f^{d,\eta ij,\eta^\prime
  kl}$, $f^{d,ij,kl}$, $f^{tn,ij,kl}$ and $f^{tn,\eta ij,\eta^\prime
  kl}$. The structure constants can be obtained from the nonvanishing
commutators as discussed above.\footnote{For $Sp(2N)$, if the
  typical off-diagonal index $b$ is denoted as $b=ij$, then $-b=ji$
  which is the same as for $SU(N)$; if $b=\eta ij$, then $-b=-\eta
  ij$.} As before, these structure constants can be obtained also by
the trace calculation. With the definition of the structure constants,
it is very straightforward to write down the equations corresponding to
Eqs.~(\ref{deff}) and~(\ref{deff2}).

The background field of $Sp(2N)$ can be parametrized as
$Q=\s_3\otimes Q^\prime$ where $Q^\prime={\rm
  diag}(q_1,q_2,\cdots,q_N)$ is an $N\times N$ diagonal matrix. Using
the commutators
\ba
[Q, E^{ij}]&=&(q_i-q_j)E^{ij}\, ,\nonumber\\{}
[Q, E^{\eta ij}]&=&\eta(q_i+q_j)E^{\eta ij}\, ,
\ea
we found that the arguments of the Bernoulli functions are the
following: $q_d=0$, $q_{ij}=q_i-q_j$ and $q_{\eta ij}=\eta(q_i+q_j)$.

For the $N$ independent variables $q_i$ of the background field, with
the same assumption on their values as $SO(2N+1)$ and $SO(2N)$, we can
compute the effective potential for $Sp(2N)$ with our program. For
example, with $N=2$, the results for $Sp(4)$ are given by
\ba
\Gamma^{(2)}_f&=&\frac{5 g^2 T^4}{24}\biggl[  \frac{27}{2} q_2^4 - 10 q_2^3 + q_2 (2 q_1 -2 q_1^2-1) +
 q_2^2 (\frac{9}{2} -7q_1 + 9 q_1^2) + \frac{q_1}{2}\nonumber \\
  && ( 17q_1 -34q_1^2 + 27 q_1^3-4)\biggr]\, , \nonumber \\
\Gamma^{(2)}_i&=& g^2 T^4\biggl[30 q_2^3 - 36 q_2^4 - 2 (1 - 3 q_1)^2 q_1 (2 q_1-1) + q_2 (1 -2q_1 + 2q_1^2) -
 2 q_2^2 \nonumber \\ &&(6 -11q_1 + 12 q_1^2)\biggr]\, .
\la{sp42loop}
\ea
The effective one-loop potential is
\be
\Gamma^{(1)}=-\frac{2 \pi^2 T^4 }{9}+\frac{8 T^4 \pi^2}{3} \biggl[ 9
  q_2^4 -8 q_2^3 + q_2^2 (3 -6q_1 + 6q_1^2) + q_1^2 (3 -10q_1 + 9
  q_1^2) \biggr]\, .
\la{sp41loop}
\ee
Therefore,
\be
\frac{\Gamma^{(2)}_f+\Gamma^{(2)}_i}{\Gamma^{(1)}}=-\frac{15g^2}{16
  \pi^2}\, ,\,\,\,\,\, {\rm for}\,\, N=2\,.
\ee
For $Sp(2N)$, we verify Eq.~(\ref{twoloopresult}) explicitly up to $N=5$.

\subsection{Calculation for $G(2)$}
\la{g2}

\begin{figure}[H]
\begin{center}
\includegraphics[width=8cm]{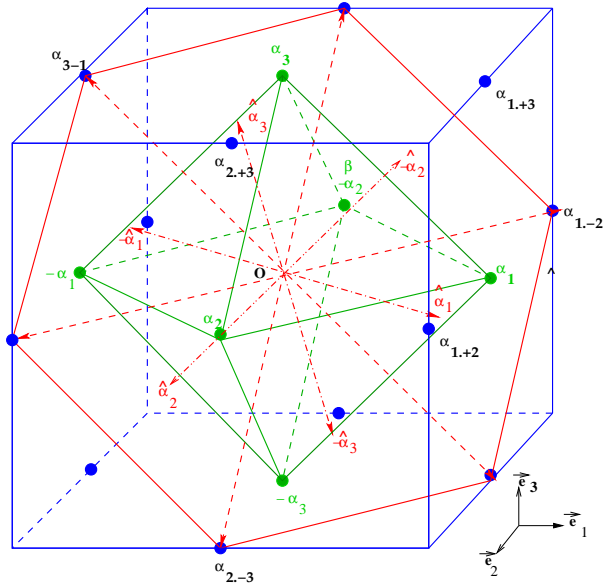}
\end{center}
\caption{The three-dimensional root space of $SO(7)$, with the plane
  where the six roots ${\vec\a}^{\,\eta i.\eta^\prime j}$
  lie. This plane is the root space of $G(2)$, on which the six short
  roots ${\vec\a}^{\,\eta i}$ of $SO(7)$ are projected. The six
  projections ${\vec\a}^{\,\eta i}$ are of length ${1\over{\sqrt{3}}}$
  in units of the length of the six roots ${\vec\a}^{\,\eta
    i.\eta^\prime j}$.}  \la{fig:so7g2project}
\end{figure}
$G(2)$ is a subgroup of $SO(7)$. It leaves the structure constants
of the octonians invariant, and this is the way it is traditionally
defined.  However, the algebra of $G(2)$ is related in a
straightforward way, by simple projections, to that of $SO(7)$ as
shown in Fig.~(\ref{fig:so7g2project}). In what follows the indices
$i$,$j$ run from $1$ to $3$ and $i>j$. The relation between the two
groups is quite simple: six of the twelve long roots ${\vec\a}^{\,\eta
  i.\eta^\prime j}$ are in the plane $q_1+q_2+q_3=0$ as shown in
Fig.~(\ref{fig:g2rootsq123}). They are the roots associated with the
generators $E^{+i.-j}$ and $E^{-i.+j}$. The other six that are not
in that plane are projected onto that plane. They are the
projections of the six short roots of $SO(7)$ which are associated
with the generators $E^{\eta i}$.

\begin{figure}[htb]
\begin{center}
\includegraphics[width=10cm]{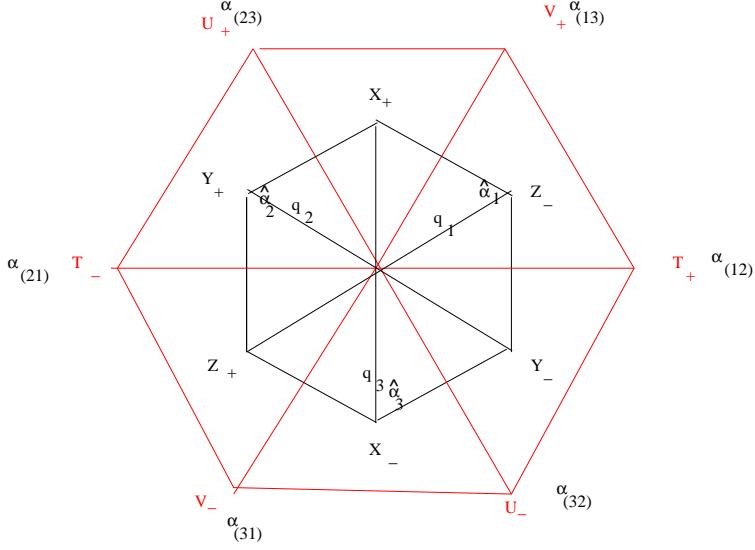}
\end{center}
\caption{The root space of $G(2)$, which is the $q_1+q_2+q_3=0$ plane
  in Fig.~(\ref{fig:so7g2project}), with the same notation for the
  roots. The roots are related to the matrices $T^{\pm}$
  etc.\ in Ref.~\cite{Holland:2003jy}.}  \la{fig:g2rootsq123}
\end{figure}
Because of the projection, the short roots of $G(2)$ are
${1\over{\sqrt{3}}}$ in units of the long roots. Recall that the short
roots of $SO(7)$ were ${1\over{\sqrt{2}}}$ in units of the long roots.
The projection respects the commutation relations of $SO(7)$, except
for the scale factor we just mentioned. For example, in $SO(7)$, we
have
\be
[E^{\eta i},E^{\eta j}]={i\over 2}\; E^{\eta i.\eta j}\, .
\ee
The generator on the right-hand side projects onto $-\eta
\epsilon_{ijk}E^{-\eta k}$. As a result, for $G(2)$, the above
commutator reads
\be
[E^{\eta i},E^{\eta j}]=-\frac{i\eta}{\sqrt{3}}\epsilon_{ijk} E^{-\eta k}\, .
\la{newcom}
\ee

The three commuting generators of the $SO(7)$ Cartan algebra reduce to
two for $G(2)$, because of the constraint $q_1+q_2+q_3=0$. For $G(2)$,
we define the two Cartan generators
\ba
H^1&=&{1\over{\sqrt{2}}}(M^{12}-M^{34})\, ,\\
H^2&=&{1\over{\sqrt{6}}}(M^{12}+M^{34}-2M^{56})\,.
\la{g2cartan}
\ea
The prefactor ensures that $\Tr (H^d)^2=\frac{1}{2}$.  Together
with the other twelve off-diagonal generators $E^{+i.-j}$, $E^{-i.+j}$
and $E^{\eta i}$, we have the explicit form of all the $14$ generators
for $G(2)$. Except for the one given by Eq.~(\ref{newcom}), all other
commutation relations can be obtained from the corresponding equations
of $SO(7)$, i.e.\ Eqs.~(\ref{3com}) and~(\ref{4com}). In addition, the
commutation relations involving the diagonal generators are
\ba
[H^{1},E^{\eta i}]&=&\frac{\eta}{2\sqrt{2}}(\d_{1i}-\d_{2i})E^{\eta i}\, ,\nonumber\\{}
[H^{2},E^{\eta i}]&=&\frac{\eta}{2\sqrt{6}}(\d_{1i}+\d_{2i}-2\d_{3i})E^{\eta i}\, ,\nonumber\\{}
[H^{1},E^{\eta i.\eta^\prime j}]&=&\frac{1}{2\sqrt{2}}\bigg(\eta(\d_{1i}-\d_{2i})+\eta^\prime(\d_{1j}-\d_{2j})\bigg)
E^{\eta i.\eta^\prime j}\, ,\nonumber\\{}
[H^{2},E^{\eta i.\eta^\prime j}]&=&\frac{1}{2\sqrt{6}}\bigg(\eta(\d_{1i}+\d_{2i}-2\d_{3i})+
\eta^\prime(\d_{1j}+\d_{2j}-2\d_{3j})\bigg)
E^{\eta i.\eta^\prime j}\, .
\ea
The above commutators state that the square of the roots ${\vec
  \a}^{\,\eta i.\eta^\prime j}$, which are associated with the $6$
generators $E^{\eta i.\eta^\prime j}$, equal $1/2$ and the square of
the roots ${\vec \a}^{\,\eta i}$ associated with the $6$ generators
$E^{\eta i}$ equal $1/6$. Since the rank of $G(2)$ is $2$, we get
$C_2(A)=2$.

We can obtain all the structure
constants that are needed to compute the effective potential through
\ba
f^{d,\eta i,\rho j}&=&2 {\rm Tr}(E^{\rho j}\cdot [H^d,E^{\eta j}])\nonumber\\{}
f^{d,\eta j.-\eta k,\rho l.-\rho m}&=&2 {\rm Tr}(E^{\rho l.-\rho m}\cdot[H^d,E^{\eta j.-\eta k}])\,,\nonumber\\{}
f^{\eta i.-\eta j,\rho k,-\rho l}&=&2{\rm Tr}( E^{-\rho l}\cdot[E^{\eta i.-\eta j},E^{\rho k}])\,,\nonumber\\{}
f^{\eta i.-\eta j,\rho k.-\rho l,\sigma t.-\sigma n}&=&2{\rm Tr}(E^{\sigma t.-\sigma n}\cdot[E^{\eta i.-\eta j},E^{\rho k.-\rho l}])\,.
\ea
In addition, we have a special one from Eq.~(\ref{newcom}) which is
$f^{\eta i,\eta j,\rho k}=-\frac{i  \eta}{\sqrt{3}}\epsilon_{ijk}\d_{\eta\rho}$.
We employ these expressions to compute the structure constants in our
\textsc{Mathematica} program.

The background field of $G(2)$ can be parametrized in the same way as
$SO(7)$, with an additional constraint that $q_3=-q_1-q_2$. Furthermore, for
the possible values of $q_i$, we use the same assumptions as for
$SU(3)$. As a result, the constraint becomes $q_3=-q_1-q_2+n$ with
$n=0,1,2$. Notice that for $G(2)$, the argument of the Bernoulli
functions can be $0$, $\eta q_i$ and $\eta (q_i-q_j)$. Unlike $SO(7)$,
there is no $q_i+q_j$ in the Bernoulli functions and our assumption on
the values of $q_i$ enables us to avoid the modulo function and also
make the sign function definitive.

The resulting effective potential for $G(2)$ reads
\ba
\Gamma^{(2)}_f&=&\frac{g^2 T^4}{3}\biggl[\frac{7}{12} +  q_3 -  q_2 +
   4 q_1^4 + 4 q_3^2 (1 + q_3 + q_3^2) -
      q_3 (1 + q_3) (1 + 6 q_3) q_2 +\nonumber \\ && (4 + 5 q_3 + 10 q_3^2) q_2^2 -
      2 (1 + 3 q_3) q_2^3 + 4 q_2^4 - 2 q_1^3 (4 + 3 q_3 + 3 q_2) +
      q_1^2 (7 + 8 q_3 + \nonumber \\ &&10 q_3^2 + 5 q_2 + 10 q_2^2) -
      q_1 (3 + 4 q_3+ 10 q_3^2+ 6 q_3^3+q_2+ 7 q_2^2+6 q_2^3 )\biggr]\, , \nonumber \\
\Gamma^{(2)}_i&=& \frac{g^2 T^4}{9}\biggl[-32 q_1^4 - q_3(3 + 8 q_3) (2 + 3q_3 + 4 q_3^2) +
   q_3 (25 + 57q_3 + 52 q_3^2) q_2 -\nonumber \\ && (25 + 69 q_3 + 72 q_3^2) q_2^2 +
   4 (3 + 13 q_3) q_2^3 - 32 q_2^4 + q_1^3 (60 + 52 q_3 + 52 q_2) -\nonumber \\ &&
   q_1^2 (34 + 78 q_3 + 72 q_3^2 + 69 q_2 + 72 q_2^2) +
   q_1 (6 + 34 q_3 + 66q_3^2 + 52 q_3^3 +\nonumber \\ && 25 q_2 + 57q_2^2 + 52 q_2^3)\biggr]\, .
\la{g22loop}
\ea
Comparing to the one-loop result
\ba
\Gamma^{(1)}&=&-\frac{14 \pi^2 T^4 }{45}+\frac{4 T^4 \pi^2}{3} \biggl[(q_1-1)^2 q_1^2 + (q_1-1 - q_3)^2 (q_1 - q_3)^2 + (q_3-1)^2 q_3^2 \nonumber \\ &&+ (q_1-1 -
     q_2)^2 (q_1 - q_2)^2 + (q_3 - q_2)^2 (1 + q_3 - q_2)^2 + (q_2-1)^2 q_2^2 \biggr]\, ,
\la{g21loop}
\ea
we see that
\be
\frac{\Gamma^{(2)}_f+\Gamma^{(2)}_i}{\Gamma^{(1)}}=-\frac{5g^2}{8 \pi^2}\,.
\ee
However, unlike for $SU(N)$, to obtain this result we must explicitly
use that $q_3=-q_1-q_2+1$ or $q_3=-q_1-q_2+2$ or $q_1=q_2=q_3=0$.

\section{A SIMPLIFIED FORM FOR THE INSERTION}\la{sim}

The insertion diagram involves sums over diagonal indices $d$ which
can be performed quite easily as they correspond to inner products
between the corresponding roots. We use the relation between the
roots and the unit vectors mentioned in the previous sections to
reduce the inner products to sums of Kronecker $\d$'s. In addition,
the antisymmetry of the Bernoulli polynomials $\widehat B_1$ and
$\widehat B_3$ is needed. For example, $\widehat B_1(\eta q_i+\eta'
q_j)=\eta' \widehat B_1(\eta\eta'q_i+q_j)$, etc. Those are then
applied to the expression for $\G_i^{(2)}$ from
Eq.~(\ref{newoneloopinsertfinal1}).

For $SU(N)$ this is quite simple. Using
\be
\la{resu3}
{\vec \a}^{\,ij}\widehat B_n(q_i-q_j)=\sqrt{2}\;
{\vec e}_i\; \widehat B_n(q_i-q_j)\,,
\ee
we get
\be
\G_i^{(2)}(SU(N))=4 g^2\sum_{ijl}\widehat B_1(q_i-q_j)\widehat B_3(q_i-q_l)\,.
\ee
Here, the only constraint on the indices is that $i\neq j$ and $i\neq
l$. In Eq.~(\ref{resu3}), $\widehat B_n$ always refers to $\widehat
B_1$ or $\widehat B_3$ and this applies throughout this section.

For the orthogonal groups the long and short roots satisfy the
following relations\footnote{On the left-hand side of
  Eq.~(\ref{reso2n}), $j>i$ is forbidden according to our
  notations. However, on the right-hand side of this equation, $j>i$
  is permitted. The same is true for Eq.~(\ref{resp2n}).}
\ba
\la{reso2n}
\sum_{\eta\eta^\prime}{\vec \a}^{\,\eta i.\eta^\prime j}\widehat B_n(\eta q_i+\eta^\prime q_j)&=&{\vec e}_i\bigg(\widehat B_n(q_i+q_j)+\widehat B_n(q_i-q_j)\bigg)\, , \\
\sum_{\eta}{\vec \a}^{\,\eta i}\widehat B_n(\eta q_i)&=&{\vec
  e}_i\widehat B_n(q_i)\,.
\ea
As a result, the insertion diagram for $SO(2N)$ is reduced to
\be
\G_i^{(2)}(SO(2N))=2 g^2\sum_{i,j,l}\bigg(\widehat B_1(q_i+q_j)+
\widehat B_1(q_i-q_j)\bigg) \bigg(\widehat B_3(q_i+q_l)+ \widehat
B_3(q_i-q_l)\bigg)\,,
\ee
with $i\neq j$ and $i\neq l$. For $SO(2N+1)$, the result is
\ba
&\G_i^{(2)}(SO(2N+1))=\G_i^{(2)}(SO(2N))+2
g^2\sum_{i,j}\bigg[(\widehat B_1(q_i+q_j)+\widehat
  B_1(q_i-q_j))\nonumber \\ &\widehat B_3(q_i)+(\widehat
  B_3(q_i+q_j)+\widehat B_3(q_i-q_j))\widehat B_1(q_i)\bigg]
+2 g^2\sum_{i}\widehat B_1(q_i)\widehat B_3(q_i)\,.
\ea
and the same constraints $i\neq j$ and $i\neq l$ apply.

Finally, for $Sp(2N)$, we have
\ba
\la{resp2n}
\sum_{\eta}{\vec \a}^{\,\eta ij}\widehat B_n(\eta (q_i+q_j))&=&{\vec
  e}_i(1+\d_{ij})\widehat B_n(q_i+q_j)\, , \\
{\vec \a}^{\,ij}\widehat B_n(q_i-q_j)&=&{\vec e}_i\widehat
B_n(q_i-q_j)\,.\la{resp2n2}
\ea
In Eq.~(\ref{resp2n}), the case where $i=j$ is included. In
Eq.~(\ref{resp2n2}), $i\neq j$ applies.
The simplified insertion diagrams read
\ba
\G_i^{(2)}(Sp(2N))&=&2 g^2\sum_{i}\bigg(\sum_{j}(\widehat
B_1(q_i+q_j)+\widehat B_1(q_i-q_j))+2 \widehat B_1(2
q_i)\bigg)\nonumber \\ &\times& \bigg(\sum_{l}(\widehat
B_3(q_i+q_l)+\widehat B_3(q_i-q_l))+2 \widehat B_3(2 q_i)\bigg)\, ,
\ea
where $i\neq j$ and $i\neq l$ apply.

For $G(2)$, we can also work out the structure constants to get
\ba
\G_i^{(2)}(G(2))&=&2 g^2\sum_{ijl}\widehat B_1(q_i-q_j)\bigg(\widehat
B_3(q_l-q_j)-\widehat
B_3(q_l-q_i)\bigg)+\frac{4g^2}{3}\sum_{i}\widehat B_1(q_i)\widehat
B_3(q_i)\nonumber \\ &+&2 g^2\sum_{ij}\widehat
B_1(q_i-q_j)\bigg(2\widehat B_3(q_i-q_j)+\widehat B_3(q_i)-\widehat
B_3(q_j)\bigg)\nonumber \\ &+& \frac{2g^2}{3}\sum_{il}\widehat
B_1(q_i)\bigg(3\widehat B_3(q_i-q_l)-\widehat B_3(q_l)\bigg)\, ,
\ea
where $i>j$, $j \neq l$ and $i \neq l$.

Using the explicit expressions for $\G_i^{(2)}$ given in this section,
we can compute for larger $N$ more efficiently. At the same time, we
can easily prove that the result for $\G_i^{(2)}$ is independent of
the value of $\widehat B_1(n_0)$ for integer $n_0$. We point out that
for the last term in $G(2)$, to show the independence on $\widehat
B_1(n_0)$, we need to use the condition $q_1+q_2+q_3=0,1,2$. Notice
that although the same condition appears for $SU(N)$, it is not
actually needed to show independence of $\widehat B_1(n_0)$.

For the free energy $\G_f^{(2)}$, according to Eq.~(\ref{newuf}),
there is an obvious simplification if one of the indices in the
structure constant is diagonal. In this case, $\sum_{d}
|f^{d,b,c}|^2$ is just the square of the roots' length which is
already known. (Here, $b$ and $c$ denote off-diagonal indices.) For
instance, for $SU(N)$ such a term becomes
\be
g^2\sum_{ij} \bigg(2 \widehat B_2(0) \widehat B_2(q_i - q_j) +
(\widehat B_2(q_i - q_j))^2\bigg)\, .
\ee
It is straightforward to get these contributions for other groups and
we don't list the rest here.

For the case where $f^{a,b,c}$ has no diagonal index, there is no
obvious simplification. However, the values of these structure
constants can be simply read off from the commutators given above.

\section{CONCLUSIONS}\la{sec:conclusions}

The main result of this paper is that the two-loop renormalization of
the effective potential is very simple: the two-loop potential is
proportional to that at one loop, Eq.~(\ref{twoloopresult}). There is
nothing in the way we perform the computation that suggests such
simplicity. For $SU(N)$ groups it has long been known~\cite{Giovannangeli:2002uv,Giovannangeli:2004sg} that this
proportionality holds along the edges of the Weyl chamber.

Hence, at this order in perturbation theory the minima of the
perturbative action stay put. How this works out to three-loop order
is something that remains to be worked out.

The two-loop effective potential found here could now be supplemented
by a model for nonperturbative physics, e.g.\ along the lines of
Refs.~\cite{e-3p_cond,Dumitru:2010mj,Dumitru:2012fw}, in an attempt to
understand the {\em eigenvalue} distribution of the Polyakov loop in
the gauge theories mentioned above.

\section*{Acknowledgements}
We thank Rob Pisarski and Nan Su for useful discussions.
A.D.\ gratefully acknowledges support by the DOE Office of Nuclear
Physics through Grant No.\ DE-FG02-09ER41620 and by The City
University of New York through the PSC-CUNY Research Award Program,
Grants No. 65041-00~43 and No. 66514-00~44.
Y.G.\ gratefully acknowledges support by the European Research Council Grant No. HotLHC ERC-2001-StG-279579,
a grant from the NSFC of China with Project No.~11205035 and by Guangxi Normal
University with Project No.~2011ZD004.

\appendix

\section{BERNOULLI POLYNOMIALS}
\la{appa}

We define the Bernoulli polynomials,
\ba
\widehat B_{d-2k}(x)&=&T\sum_{n_0}\int {d^{d-1}\vec
  p\over{(2\pi)^{d-1}}} {1\over{(p^{ij})^{2k}}}\,,\\
\widehat B_{d-2k+1}(x)&=&T\sum_{n_0}\int {d^{d-1}\vec
  p\over{(2\pi)^{d-1}}} {p_0^{ij}\over{(p^{ij})^{2k}}}\,,\\
\widehat B_d(x)&=&T\sum_{n_0}\int {d^{d-1}\vec p\over{(2\pi)^{d-1}}}
\bigg(\log (p^{ij})^2 - \log p^2\bigg)\, .
\la{bernoulli}
\ea
In these equations, $p_0^{ij}=2\pi T (n_0 + x^{ij})$ with $n_0$ an
integer. Below, the indices $i$ and $j$ associated with $x$ are
omitted for simplicity of notation. Also, $(p^{ij})^2=(p_0^{ij})^2+{\vec
  p}^{\,2}$ and $p^2=(2\pi n_0 T)^2+{\vec p}^{\,2}$.

In $d=4$ dimensions and for $k=0$, 1, we have the following four
Bernoulli polynomials:
\ba
\widehat B_4(x)&=&{2\over 3}\pi^2T^4 B_4(x)\, ,\nonumber\\
\widehat B_3(x)&=&{2\over 3}\pi T^3 B_3(x)\, ,\nonumber\\
\widehat B_2(x)&=&{1\over 2}T^2 B_2(x)\, ,\nonumber\\
\widehat B_1(x)&=&-{T\over{4\pi}} B_1(x)\, ,
\la{bernoulli2}
\ea
with
\ba
B_4(x)&=&x^2(1-x)^2\, ,\nonumber\\
B_3(x)&=&x^3-{3\over 2}x^2+{1\over 2}x\, ,\nonumber\\
B_2(x)&=&x^2-x+{1\over 6}\, ,\nonumber\\
B_1(x)&=&x-{1\over 2}\, .
\la{bernoulli3}
\ea
The above expressions are defined on the interval $0\le x \le1$ and
they are periodic functions of $x$, with period $1$. For arbitrary
values of $x$, the argument of the above Bernoulli polynomials should
be understood as $x-[x]$ with $[x]$ the largest integer less than or
equal to $x$, which is nothing but the modulo function.

If $-1\le x \le1$ we can drop the modulo functions
and the Bernoulli polynomials reduce to
\ba
B_4(x)&=&x^2(1-\epsilon(x)x)^2\, ,\nonumber\\
B_3(x)&=&x^3-{3\over 2}\epsilon(x)x^2+{1\over 2}x\, ,\nonumber\\
B_2(x)&=&x^2-\epsilon(x)x+{1\over 6}\, ,\nonumber\\
B_1(x)&=&x-{1\over 2}\epsilon(x)\, ,
\la{bernoulli4}
\ea
where $\epsilon(x)$ is the sign function.

In fact the Bernoulli polynomials $B_1(x)$ and $B_3(x)$ are odd functions of $x$, while $B_2(x)$ and $B_4(x)$ are even
functions of $x$, so we can always make the arguments of Bernoulli polynomials
positive(or be zero) and ignore the sign functions which can save a lot of
computing time. However, we point out that $B_1(x)$ has discontinuities at integer $x$. For
example, the value of $B_1(0)$ depends on the way one approaches zero,
from above or from below. If the result of the effective potential depends on $B_1(0)$, we have to know
how one approaches zero in order to use the correct values of $B_1(0)$. In this case, the sign function in
$B_1(x)$ is very important and can not be dropped even $x\ge 0$.
Fortunately, we can prove that the contributions
related to $B_1(n_0)$ vanish without specifying the value of
$B_1(n_0)$. Therefore, the effective potential does not depend on
$B_1(n_0)$ and we can simply drop the sign functions when $x\ge 0$. The proof is straightforward when using the
total antisymmetry of the structure constants. Alternatively, one can
also prove it by using the simplified expressions of the insertion
diagram given in Sec.~\ref{sim}.

\section{THE IDENTITY EQUATION~(\ref{identitytwoloop})}
\la{appb}

This identity relates the one-gluon correction of the multiply
winding Polyakov loop to the one-gluon correction of the Polyakov loop
with single winding,
\be
\<\Tr(\overline{Q_0^2}\cdot({\bf L}^n)'')\>
=\<\overline{Q_0^2}\cdot{\bf L}''\>^d~t^n_d(\Phi)\,.
\la{factortld1}
\ee
Here, $D^{ii}=\sum_j \lambda^{ij}\lambda^{ji}$ and the definition of
$\Delta_{(r)}$ is\footnote{The dependence on the argument $\Phi_{ij}$
  is through $p^{ij}$. We have $(p^{ij})^2=(p_0^{ij})^2+{\vec
    p}^{\,2}$ and $p_0^{ij}=2\pi T n_0 + T \Phi_{ij}$.}
\be
 \Delta_{(r)}(\Phi_{ij})= \sumint_p{1\over{(p_0^{ij})^r}}\Delta_{00}(p^{ij})\,,
 \la{deltarfunctions}
\ee
with
\be
\Delta_{00}(p^{ij}) = {\delta_{00}-(1-\xi) {(p_0^{ij})^2 \over
    {(p^{ij})^2}} \over{(p^{ij})^2}}\,.
\ee

The multiply winding loop $\Tr({\bf L}(A_0)^n)$ can be written as a time-ordered product from time $\tau=0$ to $\tau={n\over T}$,
\be
\Tr ({\bf L}(A_0)^n) = \Tr{\cal P} \exp\left(
i\int _0^{n/T}d\tau A_0(\vec x,\tau)\right)\,.
\la{timeorderedPl}
\ee
There is a caveat: the field $A_0$ is periodic modulo ${1\over T}$,
{\it not} ${n\over T}$. This smaller periodicity is guaranteed by the
Matsubara frequencies being integer multiples of $2\pi T$.  The propagator
follows from the action and has the small periodicity $1/T$.

Diagonal gluons do not feel the background field; upon integration
over the emission and absorption times they are odd in the Matsubara
frequency and so do not contribute. We only have to consider the
contractions of off-diagonal $\<Q_0^{ij}(\tau_2)Q_0^{ji}(\tau_1)\>$.
These propagators are gauge field propagators in $\xi$ gauge,
\be
\<Q_0^{ij}(\tau_2)Q_0^{ji}(\tau_1)\>=\exp(ip_0(\tau_2-\tau_1))\;
\Delta_{00}(p^{ij})\,.
\la{propagator00}
\ee

Note the shift of the Matsubara frequencies in the propagator
$\Delta_{00}(p^{ij})$. This follows from the diagonalization in color
space of the bilinear part of the action. The propagators
in~(\ref{propagator00}) are still periodic modulo $1/T$.

Thus, the calculation of the one-loop average of $\Tr ({\bf L}(A_0)^n)$
boils down to the one-gluon exchange correction in
\be
\Tr\overline{{\cal P}\exp\left(i\int_0^{n/T}A_0(\vec x,\tau)\right)}\,.
\ee
For convenience in what follows we write $\Phi$ instead of ${\bf
  q}=\Phi/2\pi$ in the arguments of ${\bf L}$ and the exponents. The
calculation of the average is now quickly achieved,
\ba
\left<\Tr\overline{{\cal P}\exp(i\int_0^{n/T}A_0(\vec x,\tau))}\right>&=&
-g^2\int_0^{n/T} \int_0^{\tau_1}d\tau_1d\tau_2 \Bigg<\Tr\exp(i\Phi
T\tau_2) Q_0(\tau_2)\nonumber\\
&\times&
\exp(i\Phi T (\tau_1-\tau_2)) Q_0(\tau_1) \exp(i\Phi
(n-T\tau_1))\Bigg>\,.
\la{orderedlTT}
\ea
We now use the propagator~(\ref{propagator00}) for
$\<Q^{ij}_0(\tau_2)Q^{ji}_0(\tau_1)\>$  and the identities
\be
\exp(i\Phi T\tau)\lambda_{ij}\exp(-i\Phi T\tau)=\exp(i\Phi_{ij}T\tau)\lambda_{ij}\,,
\ee
to shift the Matsubara frequencies from $p_0$ to
$p_0+\Phi_{ij}T=p_0^{ij}$.  The result is that we can drop the heavy
quark propagators but shift the frequency $p_0$ in the propagator in
Eq.~(\ref{propagator00}) to $p_0^{ij}=p_0+\Phi_{ij}T$.

The time-ordered integrals give two terms,
\be
 \int_0^{n/T}\int_0^{\tau_1}d\tau_1d\tau_2\exp(ip_0^{ij}(\tau_2-\tau_1))=
     {(1-\exp(-in\Phi_{ij})) \over{(p_0^{ij})^2}}- {n\over {iTp_0^{ij}}}\,.
\la{orderedlT}
\ee
We do the same for the mode with
$\<Q^{ji}_0(\tau_2)Q_0^{ij}(\tau_1)\>$.  As expected it gives
Eq.~(\ref{orderedlT}) with $i\leftrightarrow j$. If we sum over $-p_0$
instead of $p_0$, we see that the denominator of the first resp.\ second
term are even resp.\ odd under interchange of $i$ and $j$.
Substituting into Eq.~(\ref{orderedlTT}) we get the combination
(remember that $D^{ii}=\lambda^{ij}\lambda^{ji}$)
\ba
\<\Tr(\overline{Q_0^2}\cdot({\bf L}^n)'')\>&=&-{g^2\over 2}\sum_{ij}\Tr\bigg[{n\over {iT}}\exp(in\Phi)(D_{ii}-D_{jj})\Delta_{(1)}(\Phi_{ij})\nonumber\\
&+&\exp(in\Phi)\Delta_{(2)}(\Phi_{ij})\bigg(D^{ii}(1-\exp(-inT\Phi_{ij}))\nonumber\\
&+&D^{jj}(1-\exp(inT\Phi_{ij}))\bigg)\bigg]\,.
\ea

The second term, proportional to $\Delta_{(2)}$ drops out after taking
the trace. The reason is that $\Tr(\exp(in\Phi)D^{ii})={1\over
  2}\exp(in\Phi_i)$. Clearly, the untraced loop contains unphysical
results like the divergent $\Delta_{(2)}$ but the trace projects
them out.

The latter argument is not only valid for $SU(N)$ but also for the
other classic groups (by using the roots $\vec e_i$).


Remarkably, the matrix $t^n_d(\Phi)$ factors out and we obtain
Eq.~(\ref{identitytwoloop}),
\be
\<\Tr(\overline{Q_0^2}\cdot({\bf L}^n)'')\>=\<\overline{Q_0^2}\cdot{\bf L}''\>_d~t^n_d(\Phi)\,.
\la{factortld}
\ee
This the desired factorization, and the first
factor $\<\overline{Q_0^2}\cdot{\bf L}''\>_d$ is the projection on $\l_d$
of the one-gluon corrected Polyakov loop.

Recall that the insertion diagram in Fig.~\ref{fig:insertion} involves
only summation over the looping index $n$. The derivative acting
on the gauge field vertices is, according to~(\ref{hatzeromomn}),
\be
 \widehat Q_0^{n} \equiv \sum_d~t^n_d({\bf q})\overline{\overline{Q_0^d}}\,.
 \ee
Hence, the summation over these  loop indices $n$  drops out, because of
the factorization we just obtained,
\ba
\G_i&=&\sum_{n=1}^{N-1}\<\Tr(\overline{Q_0^2}\cdot({\bf
  L}^n)'')\> \left<{\partial S_{int}\over{\partial
{\widehat Q^n_0 }}}\right>\nonumber\\
&=&\sum_{d=1}^{N-1}\<\overline{Q_0^2}\cdot{\bf L}''\>_d \left<{\partial
  S_{int}\over{\partial
{\overline{\overline{Q_0^d}}}}}\right>\,.\\
\<\overline{Q_0^2}\cdot{\bf L}''\>_d&=&g^2{1\over{2T}}\sum_{ij}
f^{d,ij,ji}\; \Delta_{(1)}(\Phi^{ij})\,.
\la{oneloopinsertalmostfinal}
\ea
This is the result for all covariant background gauges and all
classical groups. We leave it to the reader to isolate the part
proportional to the gauge parameter $\xi$. This follows immediately
from the expression for $\Delta_{(1)}$ in Eq.~(\ref{deltarfunctions}).



\end{document}